\documentclass[preprint]{elsarticle}

\makeatletter
\def\ps@pprintTitle{%
 \let\@oddhead\@empty
 \let\@evenhead\@empty
 \def\@oddfoot{\leftline{This paper is accepted for publication in \emph{Applied Energy} (2019).}}%
 \let\@evenfoot\@oddfoot}
\makeatother

\usepackage{lineno,hyperref}
\modulolinenumbers[5]

\journal{Applied Energy}

\usepackage{color}

\usepackage{graphicx}
\graphicspath{{figs_and_tables/}}
\DeclareGraphicsExtensions{.pdf,.jpeg,.png,.jpg,.eps,.PNG,.bmp}
\usepackage{subfig}
\usepackage{multirow}
\newcommand\Tstrut{\rule{0pt}{1.0\normalbaselineskip}}         
\newcommand\Bstrut{\rule[-1.4ex]{0pt}{1.0\normalbaselineskip}}   

\usepackage{gensymb}
\usepackage{amsmath}
\DeclareMathOperator*{\argmin}{arg\,min}

\usepackage{algorithm}
\usepackage[]{algpseudocode}

\usepackage[inline]{enumitem}

\usepackage[]{nomencl}
\makenomenclature
\usepackage{etoolbox}
\renewcommand\nomgroup[1]{%
  \item[\bfseries
  \ifstrequal{#1}{A}{Acronyms}{%
  \ifstrequal{#1}{V}{Variables}}%
]}
\newcommand{\nomunit}[1]{%
\renewcommand{\nomentryend}{\hspace*{\fill}#1}}
\usepackage{framed} 
\usepackage{longtable}
\usepackage{mdframed}

\begin{document}

\begin{frontmatter}

\title{Benchmarking air-conditioning energy performance of residential rooms based on regression and clustering techniques\tnoteref{t1}}
\tnotetext[t1]{Published version in \emph{Applied Energy}: https://doi.org/10.1016/j.apenergy.2019.113548}

\author[sutd]{Yuren~Zhou\corref{cor1}}
\ead{yuren\_zhou@mymail.sutd.edu.sg}
\cortext[cor1]{Corresponding author}

\author[sutd]{Clement~Lork}
\ead{clement\_lork@mymail.sutd.edu.sg}

\author[sutd]{Wen-Tai Li}
\ead{wentai\_li@sutd.edu.sg}

\author[sutd]{Chau~Yuen}
\ead{yuenchau@sutd.edu.sg}

\author[bcaa]{Yeong~Ming~Keow}
\ead{keowym@gmail.com}

\address[sutd]{Singapore University of Technology and Design, 8 Somapah Road, Singapore}
\address[bcaa]{Building and Construction Authority Academy, 200 Braddell Road, Singapore}

\begin{abstract}

Air conditioning (AC) accounts for a critical portion of the global energy consumption. To improve its energy performance, it is important to fairly benchmark its energy performance and provide the evaluation feedback to users.
However, this task has not been well tackled in the residential sector.
In this paper, we propose a data-driven approach to fairly benchmark the AC energy performance of residential rooms.
First, a regression model is built for each benchmarked room so that its power consumption can be predicted given different weather conditions and AC settings. Then, all the rooms are clustered based on their areas and usual AC temperature set points. Lastly, within each cluster, rooms are benchmarked based on their predicted power consumption under uniform weather conditions and AC settings.
A real-world case study was conducted with data collected from 44 residential rooms. Results show that the constructed regression models have an average prediction accuracy of 85.1\% in cross-validation tests, and support vector regression with Gaussian kernel is the overall most suitable model structure for building the regression model.
In the clustering step, 44 rooms are successfully clustered into seven clusters.
By comparing the benchmarking scores generated by the proposed approach with two sets of scores computed from historical power consumption data, we demonstrate that the proposed approach is able to eliminate the influences of room areas, weather conditions, and AC settings on the benchmarking results. Therefore, the proposed benchmarking approach is valid and fair.
As a by-product, the approach is also shown to be useful to investigate how room areas, weather conditions, and AC settings affect the AC power consumption of rooms in real life.

\end{abstract}


\end{frontmatter}


\section{Introduction}
\label{sec_intro}


Buildings are now consuming over half of the global electricity~\cite{IEABuilding2015} and 18.5\% of the consumption is used for space cooling~\cite{IEACooling2018}. In hot and humid countries such as Singapore, this portion can become as high as 50\%~\cite{NEA2010}. Meanwhile, the demands of air conditioning (AC) are still increasing rapidly due to global warming~\cite{isaac2009modeling, wong2010impact}, and supplemented by burgeoning demands in emerging economies such as India~\cite{IEACooling2018}. On current trends, the global energy used for AC in 2050 is predicted to be more than triple compared with 2016, with nearly 70\% of the increase coming from residential usage~\cite{IEACooling2018}. The generation of such a huge amount of electricity can cause problems such as air pollution, global warming, and heavy peak loads on power systems. Therefore, it is necessary to improve the energy efficiency of AC usage, especially in the residential sector, such that the total energy consumption can be maintained at a relatively low level even when the demand keeps expanding.
\nomenclature[A]{AC}{air conditioning}

In order to achieve better energy efficiency of AC usage, many efforts have been made to enhance the performance of hardware systems, including individual components, overall system design, and control strategies as summarized in~\cite{chua2013achieving, vakiloroaya2014review}.
While these approaches indeed improve the energy performance of new AC systems, we also need to consider the potential energy savings from existing AC systems by conducting regular system maintenance~\cite{chang2011energy}, repairing faults like envelope air leakages in the built environment~\cite{kim2013feasibility}, and improving user behaviors~\cite{pisello2014human}. Careless AC usage behaviors, such as leaving doors and windows open with AC operating or boiling water inside the air-conditioned room, need to be prevented. As a result, there is a need to raise the users' awareness of the energy performance level of their AC systems such that they can be informed on the potential leakage or poor behaviors, and adopt energy-saving strategies without the compromise of thermal comfort. To achieve this goal, benchmarking AC energy performance of different buildings or rooms and providing the results as feedback to the users is an effective approach~\cite{darby2001making}.

In current literature, existing research on energy performance benchmarking is mainly conducted for the overall energy performance of entire buildings, such as commercial buildings in~\cite{chung2012using,wang2016methodology} and multi-family housing complexes in~\cite{jeong2016development}. While some works tackle issues related to benchmarking AC energy performance specifically, the benchmarked entities are entire commercial buildings, such as office buildings in~\cite{li2017analysis} and shopping centers in~\cite{li2018benchmarking}.
To the best of our knowledge, there is little effort spent on benchmarking AC energy performance in the residential sector. Nevertheless, this is an important task.
According to the report by the International Energy Agency~\cite{IEACooling2018}, the major increase of AC energy consumption will come from the residential sector in the following years.
Unlike the commercial buildings where AC is often centrally controlled, AC in the residential buildings (e.g. apartments) is usually decentralized and controlled by users in individual rooms. Therefore, benchmarking of AC energy performance in residential buildings should be conducted in the level of rooms so that individual users can get precise feedback and take the responsibility to improve the energy performance.
Moreover, as will be shown in the following section, existing benchmarking approaches for other benchmarking targets are not applicable to this task.
As a result, a well-designed benchmarking approach for AC energy performance of residential rooms is needed, and this paper is aimed to fill this gap.

To this end, we propose a data-driven approach to benchmark AC energy performance of residential rooms, with a focus on the cooling scenario. For a set of individual rooms, given the historical AC operation data and weather condition data, the expected output of the benchmarking approach is a benchmarking score for each room, which represents the level of its AC energy performance.

In order to develop the benchmarking approach, first, three key elements, including type of benchmark, energy performance index (EPI), and noisy factors, are carefully selected based on a comprehensive review of related studies and case-specific requirements. Afterward, a procedure of conducting the benchmarking is designed so that the influences of the selected noisy factors on the benchmarking results are properly eliminated.
In the procedure, a predictive model is first built for each benchmarked room to predict its power consumption given different weather conditions and AC settings based on regression techniques. Next, all the benchmarked rooms are separated into different clusters based on their areas and usual AC temperature set points using clustering techniques. Lastly, within each cluster, rooms are benchmarked based on their predicted power consumption under uniform weather conditions and AC settings.
After proposing the benchmarking approach, a real-world case study is presented to verify the approach, using data collected from 44 rooms in an apartment building.
\nomenclature[A]{EPI}{energy performance index}

As the proposed approach is aimed at fairly benchmarking AC energy performance of residential rooms, the generated results (i.e. benchmarking scores) can tell if a given room has poor performance in peer-to-peer comparison. Residents in the room can then be motivated to take action to improve the AC energy performance in their room by getting into energy-friendly habits, conducting AC maintenance regularly, or fixing envelope air leakages.
Since the proposed approach is based on machine learning techniques, including regression and clustering techniques, it does not require any heavy-manpower work, such as manually selecting benchmarks or calibrating simulation models. At the same time, the used data types are very common and easy to collect. As a result, the proposed approach is directly applicable to other similar benchmarking applications for AC energy performance.
For benchmarking applications with different targets, such as overall energy performance of buildings, houses, or other types of systems, the proposed approach, along with the literature review conducted in this paper, can be used as a comprehensive guide of developing a proper benchmarking approach.
Although the case study is conducted based on a real-world testbed in Singapore, there is no assumption made related to the geographical position, and the proposed benchmarking approach is developed for a generally formed problem. As a result, the proposed approach can be applied worldwide, and this paper is of interest to international readers.

The contributions of this paper can thus be summarized as the following:
\begin{itemize}[noitemsep,topsep=3pt,parsep=0pt,partopsep=0pt]
\item Existing studies are reviewed and organized based on general steps of developing a benchmarking approach, which can be used as a comprehensive instruction for future research.
\item A data-driven benchmarking approach is proposed to fill the research gap of benchmarking AC energy performance of residential rooms.
\item A real-world case study is conducted using data collected from 44 rooms in an apartment building. The results verify the proposed benchmarking approach and provide experience for future research.
\end{itemize}

\noindent\rule[-12pt]{\linewidth}{0.6pt}\\[-12pt]
\printnomenclature[1.6cm]
\noindent\rule[12pt]{\linewidth}{0.6pt}

\section{Review of existing studies on energy performance benchmarking}

Energy performance benchmarking refers to evaluating the energy performance of target entities, such as the overall energy performance of entire buildings~\cite{perez2009review,wang2012quantitative}, by selecting a reasonable benchmark and comparing the energy performance index (EPI) of the entities with the benchmark. Generally, many factors affect the energy performance of the entities, and thus they are the reasons for energy performance differences between the entities. While the users should be criticized for certain reasons (e.g. poor maintenance and careless usage), there are objective factors that the users should not be blamed for (e.g. building specifications and climate). As a result, to make the benchmarking results fair, the influences of these objective factors (referred to as noisy factors in the following) on the energy performance should be eliminated before the comparison~\cite{chung2011review,wang2015benchmarking}.

\begin{figure}[b]
\centering
\includegraphics[width=.9\linewidth]{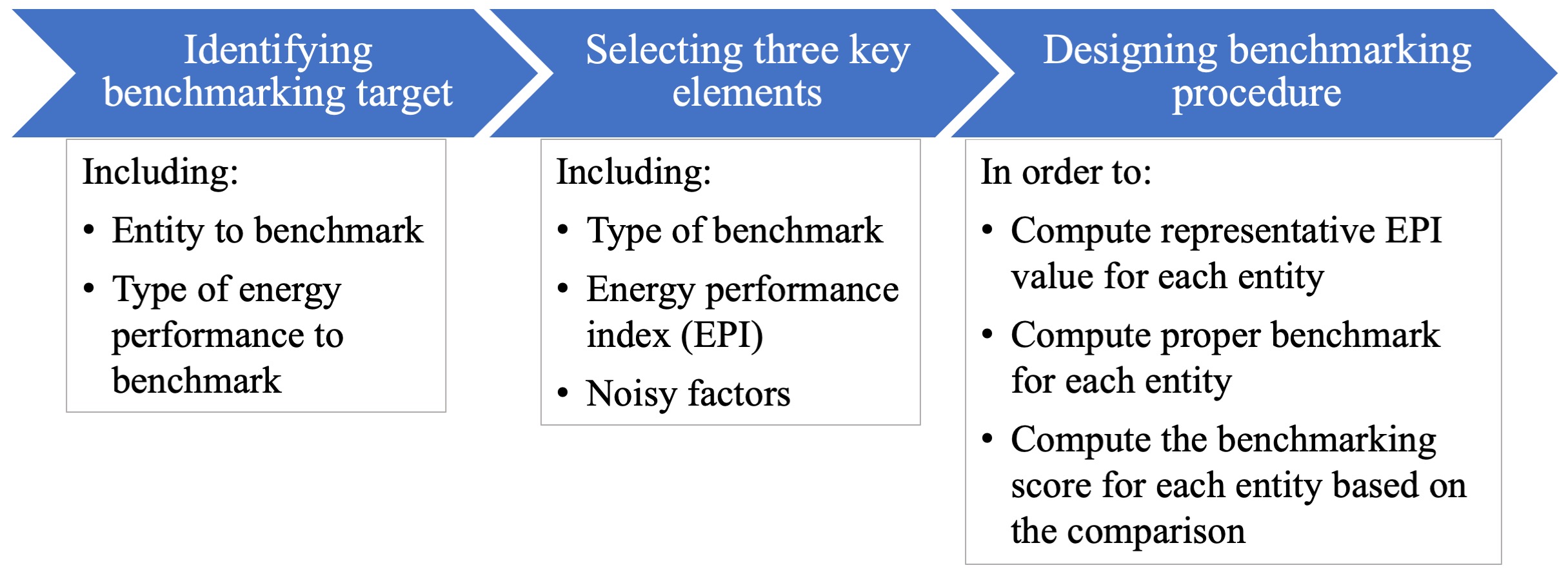}
\caption{General steps of developing a benchmarking approach.}
\label{fig_steps_develop_approach}
\end{figure}

The general steps of developing an energy performance benchmarking approach are summarized as in Fig.~\ref{fig_steps_develop_approach}. First, the benchmarking target needs to be identified clearly, including the entity and the type of energy performance to benchmark. Next, three key elements of the approach need to be selected properly, namely the type of benchmark, EPI, and noisy factors. Finally, the benchmarking procedure should be designed accordingly to accomplish three main tasks as summarized in the figure. The first two tasks are responsible for eliminating the influences of noisy factors, so as to ensure a fair comparison in the third task.

The rest of this section reviews existing studies on energy performance benchmarking, following the steps in Fig.~\ref{fig_steps_develop_approach}. For each step, common types of implementations in the existing studies are summarized, and their pros and cons are discussed.

\subsection{Review of benchmarking targets}

The first step of developing a benchmarking approach is to clearly identify the benchmarking target, including the entity and type of energy performance to benchmark, because the selection of the key elements and design of the benchmarking procedure all depend on this. In another word, when the benchmarking target changes, a different benchmarking approach is needed.

In current literature, the most common benchmarking target is the overall energy performance of entire buildings, both commercial and residential. Examples of benchmarking the overall energy performance of commercial buildings can be found in~\cite{chung2012using} and~\cite{wang2016methodology}, where the developed benchmarking (diagnosis) approaches were applied to a supermarket and an office building in the case studies, respectively. An example of benchmarking the overall energy performance of residential buildings can be found in~\cite{jeong2016development}, where Jeong et al. benchmarked the energy efficiency of 503 multi-family housing complexes.

Other than the overall energy performance of buildings, some benchmarking studies were particularly conducted for the AC energy performance of buildings. For example, in~\cite{li2017analysis}, Li et al. analyzed the performance of a chiller system in an office building through a simulation study. In~\cite{li2018benchmarking}, Li et al. developed a benchmarking method for the cooling energy performance of large commercial buildings and verified it using data collected from eight large shopping centers.

To the best of our knowledge, there is a lack of approaches to benchmark the AC energy performance of residential rooms, whose importance has been discussed in the Section~\ref{sec_intro}. Therefore, it is necessary to develop a new benchmarking approach to fulfill this task.

\subsection{Review of three key elements}
\label{sec_review_element_selection}

This subsection reviews the three key elements of existing benchmarking approaches, namely the type of benchmarking, EPI, and noisy factors. Although existing approaches were designed for benchmarking targets other than the AC energy performance of residential rooms, understanding the pros and cons of existing choices helps to select the most suitable elements for the proposed approach.

\subsubsection{Review of types of benchmark}
\label{sec_review_type_of_benchmark}

Benchmarks can be categorized into three types based on the type of performance they represent: previous-performance benchmarks, intended-performance benchmarks, and peer-performance benchmarks~\cite{li2014methods}. Which type of benchmark to use depends on the ultimate purpose and requirements of the benchmarking application.

Previous-performance benchmarks represent the previous performance of oneself.
They can be computed by a predictive model trained with historical data of the entity or selected from historical data with certain conditions~\cite{lazarova2016fault}. This type of benchmark is useful for tracking system degradation~\cite{wang2017automated} and detecting faults~\cite{du2014fault}, but it is not able to yield meaningful benchmarking results if the average level of the previous performance is poor.

Intended-performance benchmarks represent the ideal performance that the entity should achieve by design or according to certain standards.
They can be computed from simulation tools (e.g. EnergyPlus) with the original building design as input~\cite{maile2012method} or from white-box (physics) models with real system measurements and corresponding standards as input~\cite{li2018benchmarking}.
Although this type of benchmark is the most accurate and objective, it requires heavy knowledge about the system characteristics and simulation model calibration. Hence, it is only applicable when the number of benchmarked entities is small.

Peer-performance benchmarks represent the performance of other comparable entities. On one hand, by incorporating the performance of others, the comparison is no longer limited to the best previous performance of oneself. On the other hand, unlike the intended-performance benchmarks, computing this type of benchmark is fully data-driven, so it avoids the need for specific knowledge about the system characteristics and tedious simulation model calibration for each benchmarked entity~\cite{kavousian2013data,jeong2017improvements}. In another word, this type of benchmark balances the quality of benchmarking results and the scalability of the approach.
As a result, the peer-performance benchmark is adopted in the proposed benchmarking approach due to its balanced performance. Since computing peer-performance benchmark is data-driven, the proposed approach is data-driven.

\subsubsection{Review of energy performance indices}

In existing studies on energy performance benchmarking, the simplest and most general EPI is the direct measurement of energy consumption (or equivalent $\mathrm{CO_2}$ emissions), such as the overall energy consumed by the evaluated buildings used in~\cite{kavousian2015ranking,liu2017energy}.

More often, simply-normalized EPI (i.e. energy consumption divided by one noisy factor) is used to make the number more intuitive. One typical example is the energy use intensity (EUI), which equals to energy consumption divided by building area and is commonly used in studies on building energy performance benchmarking like~\cite{chung2009study,mathew2015big}. In some studies, such as~\cite{filippin2000benchmarking,wang2017multi}, it is assumed that the simply-normalized EPI is not affected by a particular noisy factor any more after the division, and thus is used for comparison directly.
This assumption requires the energy consumption and the certain noisy factor being proportional, but this is rarely true for complex energy systems. When benchmarking building energy performance, it is still necessary to normalize the EUI by building area and other noisy factors~\cite{chung2011review}.
\nomenclature[A]{EUI}{energy use itensity}

Sometimes more complex EPIs are used to achieve more accurate evaluation, such as exact cooling load demands and cooling system efficiency used in~\cite{li2018benchmarking}. However, in order to compute such EPIs, detailed data about the energy system are necessary for every entity, which increases the cost of data collection.
For example, in~\cite{li2018benchmarking}, the computation of cooling load demands and cooling system efficiency required specifications of the building envelope, ventilation, occupants, lighting and equipment, and transmission system.
Therefore, this type of EPI is not scalable to a large number of entities.

Besides determining what type of EPI to compute, it is also important to set the correct time scope of the computation. When benchmarking the energy performance of entire buildings, the time scope of computing the EPI is usually fixed, such as average daily EPI in~\cite{kavousian2015ranking} or annual EPI in~\cite{koo2015development,park2016development}. For AC energy performance of residential rooms, however, a new computation time scope is needed, because the AC is often operated for different amount of time every day for each individual room.

From above, each type of EPI has its pros and cons in terms of easiness of understanding, cost of data collection, and accuracy as the representative of the performance. As a result, the selection of the type of EPI and its computation time scope should be conducted based on the available data resources and requirements of each particular benchmarking application. In order to develop a good benchmarking approach for the AC energy performance of residential rooms, a suitable EPI is carefully designed in Section~\ref{sec_adopted_EPI}.

\subsubsection{Review of noisy factors}

As defined at the beginning of this section, noisy factors are factors whose influences on the energy performance of the benchmarked entities are objective and should not be considered as reasons for the energy performance differences in benchmarking.
As a result, for different benchmarking targets, different sets of noisy factors should be selected.

The selection of noisy factors should be done based on the physics knowledge of the entities, as well as the availability of data that represent the noisy factors.
In studies aimed at benchmarking building energy performance, the noisy factors considered are usually building structure specifications (e.g. the number of floors, total area, building envelope properties), local weather, appliance ownership, and occupant attributes (e.g. people count and lifestyle)~\cite{kavousian2015ranking,koo2015development}.
As for benchmarking the AC energy performance of residential rooms, a thorough discussion on the system physics should be conducted before choosing the suitable set of noisy factors, which will be done in Section~\ref{sec_adopted_noisy_factors}.

\subsection{Review of benchmarking procedures}
\label{sec_review_benchmark_procedure}

The design of the benchmarking procedure should be done to fulfill the three tasks as summarized in Fig.~\ref{fig_steps_develop_approach}. Moreover, the key requirement of the first two tasks is to eliminate the influences of noisy factors. Based on how this is achieved, existing benchmarking procedures (mainly for building energy performance benchmarking) can be categorized into two types: simple normalization and noisy factor equalization.

Simple normalization adopts simply-normalized EPI for each benchmarked entity, selects the entity with the best EPI value as the benchmark, and compares the EPI of other entities with the benchmark~\cite{filippin2000benchmarking,wang2017multi}. It makes a questionable assumption that energy consumption and certain noisy factors are proportional, so it is not commonly used in recent studies.

Noisy factor equalization means computing a benchmark that shares the similar noisy factor values with the benchmarked entity, such that the comparison between the EPI value and the benchmark is fair. Existing procedures of this type mainly include simple regression analysis (SRA)~\cite{chung2006benchmarking}, stochastic frontier analysis (SFA)~\cite{buck2007potential}, data envelopment analysis (DEA)~\cite{charnes1981evaluating}, and clustering-based analysis (CBA)~\cite{gao2014new}.
\nomenclature[A]{SRA}{simple regression analysis}
\nomenclature[A]{SFA}{stochastic frontier analysis}
\nomenclature[A]{DEA}{data envelopment analysis}
\nomenclature[A]{CBA}{clustering-based analysis}

SRA and SFA first build a regression model with EPI of each entity as the dependent variable and noisy factors as the independent variables.
Next, for each entity, a benchmark is computed by inputting its values of noisy factors into the regression model. The obtained benchmark represents the average performance of all entities under the specific condition of the given entity.
SRA treats the residual between the benchmark and the actual EPI of the benchmarked entity as the inefficiency of the entity, while in fact it also contains the error of the regression model. Although SFA explicitly separates these two components by statistic approaches, the accuracy is not guaranteed~\cite{chung2011review}.

DEA and CBA compute the benchmark from the EPI values of entities that share similar noisy factor values with the benchmarked entity. By solving an optimization problem, DEA computes the benchmark that represents the best performance of the entities with similar noisy factor values and then uses the benchmark to compute the efficiency scores for all the corresponding entities~\cite{kavousian2013data}. As for CBA, all the benchmarked entities are first clustered based on their values of noisy factors. Then within each cluster, a benchmark is computed as the typical values (e.g. the mean, median, etc.) of the EPI values of all entities belong to this cluster. The benchmarking score of each entity can then be computed by comparing its EPI value with the benchmark of its cluster~\cite{park2016development}.
The main limitation of DEA and CBA is that they both require a certain amount of entities that share similar noisy factor values. When the number of noisy factors is large, the dataset of benchmarked entities needs to be very large to achieve this requirement.

Besides the drawbacks summarized above for the existing benchmarking procedures, they all have a technical deficiency if applied to benchmark AC energy performance of residential rooms: a proper way to compute the representative EPI value for every room.
On one hand, the existing procedures are mostly designed for entire buildings, and they normally compute the representative EPI value for each building by averaging the historical EPI values over a certain period of time.
On the other hand, unlike large energy systems such as entire buildings, the historical EPI values of the same residential room in different time periods can be very diverse due to different values of noisy factors (e.g. control settings). Therefore, averaging the historical EPI values is not a proper way to compute the representative EPI value for each residential room.

In summary, existing benchmarking procedures designed for other types of benchmarked entities are insufficient for benchmarking the AC energy performance of residential rooms, and thus a new benchmarking procedure is necessary, which will be proposed in Section~\ref{sec_benchmarking_procedure}.

\section{Key elements of the proposed approach}

This section describes the three key elements (i.e. type of benchmark, EPI, and noisy factors) chosen for the proposed approach for benchmarking AC energy performance of residential rooms, based on the knowledge gained from reviewing the related studies in Section~\ref{sec_review_element_selection}.

\subsection{Adopted type of benchmark}

As discussed in Section~\ref{sec_review_type_of_benchmark}, peer-performance benchmarks keep a good balance between the quality of benchmarking results and the scalability of the approach. Because the proposed approach is intended to be used for a given set of residential rooms and the number of benchmarked rooms could be large, the peer-performance benchmark is adopted.

\subsection{Adopted energy performance index}
\label{sec_adopted_EPI}

As for selecting proper EPI, due to the consideration of scalability, complex EPIs (e.g. cooling load or system efficiency) that require very detailed data types are not applicable.
Also, as will be shown in Section~\ref{sec_adopted_noisy_factors}, the energy consumption for AC and the noisy factors of a residential room are not proportional. And for a residential room, the direct measurement of energy consumption for AC is already simple and intuitive, so simply-normalized EPIs are not necessary.
As a result, the EPI adopted in the proposed approach is one of the direct AC energy measurements of the room, namely the electric power consumed by the AC system.

Moreover, considering the diversity of AC operation duration in residential rooms, the time scope of EPI computation should not be fixed as daily or annual. Instead, computing the EPI per operation segment and adding the segment duration as one noisy factor is a better choice.
As a result, the EPI of each residential room used in the proposed approach is the average power consumed by the AC system in one particular operation segment ($\bar{p}_{ac}$).

\subsection{Adopted noisy factors}
\label{sec_adopted_noisy_factors}

The precise selection of noisy factors should be done based on the physics knowledge of the power consumption of the benchmarked entity. For a residential room cooled by an individual AC system, a simplified real-time thermal dynamic model can be expressed as:
\begin{equation}
C_r \cdot \frac{dT_r}{dt} = -\dot{Q}_{ac} + \frac{1}{R_w} \cdot (T_a-T_r) + \dot{Q}_{sun} + \dot{Q}_{hum}
\end{equation}
\nomenclature[V]{$C_r$}{overall thermal capacity of the cooled room
  \nomunit{$\mathrm{J\,K^{-1}}$}}
\nomenclature[V]{$T_r$}{room temperature
    \nomunit{$\mathrm{\,^{\circ}C}$}}
\nomenclature[V]{$\dot{Q}_{ac}$}{rate of heat loss through the AC system
    \nomunit{$\mathrm{W}$}}
\nomenclature[V]{$R_w$}{thermal resistance of the wall
  \nomunit{$\mathrm{K\,W^{-1}}$}}
\nomenclature[V]{$T_a$}{ambient air temperature
    \nomunit{$\mathrm{\,^{\circ}C}$}}
\nomenclature[V]{$\dot{Q}_{sun}$}{rate of heat gain from solar radiation
    \nomunit{$\mathrm{W}$}}
\nomenclature[V]{$\dot{Q}_{hum}$}{rate of heat gain from human activities
  \nomunit{$\mathrm{W}$}}

By integrating the above equation in the time scale of an AC operation segment (one period from turning AC on to turning it off), one can get the following:
\begin{equation}
\label{equa_integrated}
C_r \cdot \Delta T_r = -Q_{ac} + \frac{1}{R_w} \cdot (\bar{T}_a-\bar{T}_r) \cdot t_{seg} + Q_{sun} + Q_{hum}
\end{equation}
\nomenclature[V]{$\Delta T_r$}{change of room temperature in an operation segment
    \nomunit{$\mathrm{\,^{\circ}C}$}}
\nomenclature[V]{$Q_{ac}$}{total heat loss through the AC system in an operation segment
    \nomunit{$\mathrm{J}$}}
\nomenclature[V]{$\bar{T}_r$}{average room temperature in an operation segment
  \nomunit{$\mathrm{\,^{\circ}C}$}}
\nomenclature[V]{$\bar{T}_a$}{average ambient temperature in an operation segment
    \nomunit{$\mathrm{\,^{\circ}C}$}}
\nomenclature[V]{$t_{seg}$}{duration of an operation segment
    \nomunit{$\mathrm{s}$}}
\nomenclature[V]{$Q_{sun}$}{heat grain from solar radiation in an operation segment
  \nomunit{$\mathrm{J}$}}
\nomenclature[V]{$Q_{hum}$}{heat gain from human activities in an operation segment
    \nomunit{$\mathrm{J}$}}

In the above equation, $C_r$ and $R_w$ can be further expressed by the room specifications as:
\begin{equation}
C_r = C_a \cdot \rho_a \cdot H_r \cdot A_r
\label{equa_Cr}
\end{equation}
\begin{equation}
R_w = \frac{U_w}{k_w \cdot A_w}
\end{equation}
\nomenclature[V]{$C_a$}{specific heat capacity of air
    \nomunit{$\mathrm{J\,kg^{-1}\,K^{-1}}$}}
\nomenclature[V]{$\rho_a$}{density of air
  \nomunit{$\mathrm{kg\,m^{-3}}$}}
\nomenclature[V]{$H_r$}{room height
    \nomunit{$\mathrm{m}$}}
\nomenclature[V]{$A_r$}{room area
    \nomunit{$\mathrm{m^2}$}}
\nomenclature[V]{$U_w$}{wall thickness
    \nomunit{$\mathrm{m}$}}
\nomenclature[V]{$k_w$}{thermal conductivity of the wall
    \nomunit{$\mathrm{W\,m^{-1}\,K^{-1}}$}}
\nomenclature[V]{$A_w$}{wall area
    \nomunit{$\mathrm{m^2}$}}

Assuming that by the end of the operation segment, the room temperature has been stabilized around the average temperature set point, then $\Delta T_r$ can be written as:
\begin{equation}
\Delta T_r = \bar{T}_{set} - T_{ri}
\end{equation}
\nomenclature[V]{$\bar{T}_{set}$}{average temperature set point in an operation segment
    \nomunit{$\mathrm{\,^{\circ}C}$}}
\nomenclature[V]{$T_{ri}$}{initial room temperature in an operation segment
    \nomunit{$\mathrm{\,^{\circ}C}$}}

$Q_{ac}$ can be computed as the total power consumption multiplied by the overall energy efficiency ratio (EER) as the following:
\begin{equation}
Q_{ac} = EER_o \cdot \bar{p}_{ac} \cdot t_{seg}
\end{equation}
\nomenclature[A]{EER}{energy efficiency ratio}
\nomenclature[V]{$EER_o$}{overall energy efficiency ratio of the AC system in a segment
    \nomunit{}}
\nomenclature[V]{$\bar{p}_{ac}$}{average power consumed by the AC system in an operation segment
    \nomunit{$\mathrm{W}$}}

Further, $Q_{sun}$ comes from the solar radiation energy with a conversion ratio which can be expressed as:
\begin{equation}
Q_{sun} = C_{s2h} \cdot \bar{p}_{si} \cdot t_{seg}
\label{equa_Qsun}
\end{equation}
\nomenclature[V]{$C_{s2h}$}{overall conversion ratio from solar radiation to heat in a segment
    \nomunit{}}
\nomenclature[V]{$\bar{p}_{si}$}{average solar irradiance in an operation segment
    \nomunit{$\mathrm{W}$}}

After substituting Equation~\ref{equa_Cr}-\ref{equa_Qsun} into Equation~\ref{equa_integrated} and adjusting the structure, the average power consumed by the AC system in an operation segment can be expressed by all the other factors as the following:
\begin{equation}
\bar{p}_{ac} = \frac{1}{EER_o} \cdot \lbrack
-\frac{1}{t_{seg}} \cdot C_a \cdot \rho_a \cdot H_r \cdot A_r \cdot (\bar{T}_{set} - T_{ri}) + \frac{k_w \cdot A_w}{U_w} \cdot (\bar{T}_a - \bar{T}_r) + C_{s2h} \cdot \bar{p}_{si} + \frac{Q_{hum}}{t_{seg}}
\rbrack
\label{equa_final}
\end{equation}

On the right-hand side of the above equation, besides the two constants $C_a$ and $\rho_a$, all the variables are factors affecting the power consumption of the AC system.
Among these, three factors are related to AC user behaviors: $EER_o$ depends on the quality and maintenance status of the AC system, $C_{s2h}$ is mostly decided by whether the curtain is pulled down inside the room, and $Q_{hum}$ comes from human activities such as leaving doors open. These three are the main reasons for the energy performance differences between AC systems, which we want to capture in benchmarking.
The rest 11 factors are noisy factors whose influences should be eliminated during the benchmarking process.

Among the 11 noisy factors, five are room specifications (i.e. $H_r$, $A_r$, $k_w$, $A_w$, and $U_w$), two are ambient weather conditions (i.e. $\bar{T}_a$ and $\bar{p}_{si}$), two are related to indoor temperature (i.e. $T_{ri}$ and $\bar{T}_r$), and two are AC operation settings (i.e. $t_{seg}$ and $\bar{T}_{set}$). Although operation settings are one type of user behavior, they are considered as noisy factors here because they represent the level of service of ACs or the desired cooling comfort by users. In another word, one should not be criticized for the reason that he or she wants a colder indoor environment.

In practice, among the five room specifications, only $A_r$ must be handled as a noisy factor, because differences in $H_r$, $U_w$, and $k_w$ between rooms in the same geographical area are often neglectable, and $A_w$ is directly proportional to $A_r$. Therefore, neglecting these four noisy factors does not affect the final results much, but reduces the complexity of the problem as well as ease the data collection process, and thus is adopted. Moreover, outdoor humidity as an important ambient weather condition also has an influence on the cooling performance of AC systems, although this is not reflected by the equations. Therefore, the average ambient relative humidity in one AC operation segment ($\bar{H}_a$) is also included in the noisy factor list.
\nomenclature[V]{$\bar{H}_{a}$}{ambient relative humidity in an operation segment
    \nomunit{$\mathrm{\%}$}}

To summarize, in total eight noisy factors are adopted in the proposed benchmarking approach for cooling energy performance of AC systems in residential rooms, including $A_r$, $\bar{T}_a$, $\bar{H}_a$, $\bar{p}_{si}$, $T_{ri}$, $\bar{T}_r$, $t_{seg}$, and $\bar{T}_{set}$.

\section{Benchmarking procedure of the proposed approach}
\label{sec_benchmarking_procedure}

\begin{figure}[!t]
\centering
\includegraphics[width=\linewidth]{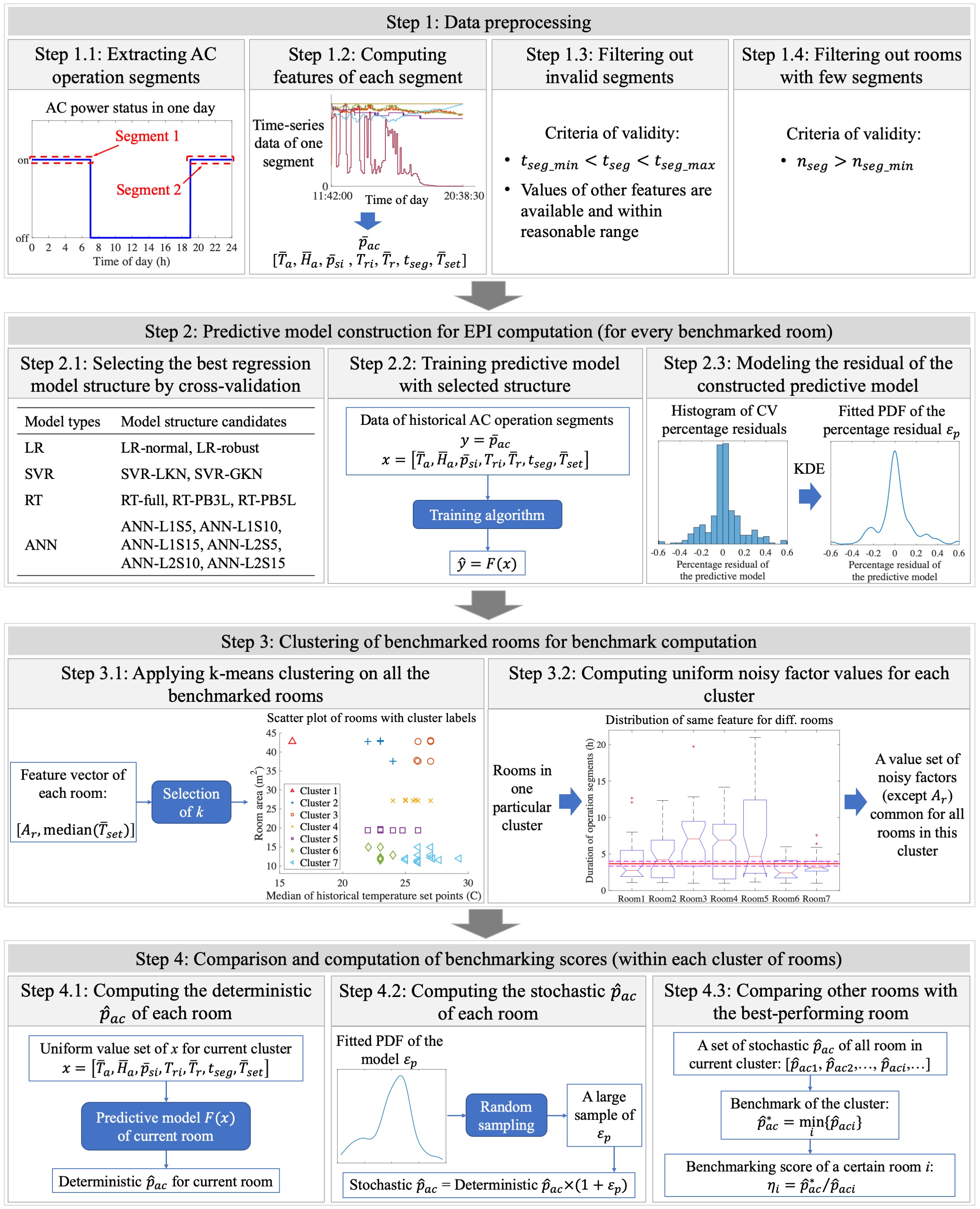}
\caption{Overall framework of the benchmarking procedure.}
\label{fig_benchmarking_framework}
\end{figure}

As summarized in Section~\ref{sec_review_benchmark_procedure}, the existing benchmarking procedures are not applicable to benchmarking the AC energy performance of residential rooms. Therefore, a new benchmarking procedure is proposed in this section, which is particularly designed for this task. The overall framework of the procedure is depicted in Fig.~\ref{fig_benchmarking_framework}, while details are described in the corresponding text.

\subsection{Step 1: Data preprocessing}
\label{sec_data_preprocess}

\begin{table}[t]
\centering
\resizebox{.82\textwidth}{!}{%
\begin{tabular}{ll}
\hline
Data categories                    & Included types                    \Tstrut\Bstrut\\
\hline
AC power consumption                        & Consumed power ($p_{ac}$)                  \Tstrut\Bstrut\\
AC operation settings      & Operation status (on or off)               \Tstrut\\ 
                                            & Operation mode (dehumidifying or cooling)  \\ 
                                            & Temperature set point ($T_{set}$)          \Bstrut\\
Indoor environment                          & Indoor temperature ($T_r$)                 \Tstrut\Bstrut\\
Room specifications                         & Room area ($A_r$)                          \Tstrut\Bstrut\\
Ambient weather conditions & Outdoor temperature ($T_{a}$)              \Tstrut\\ 
& Outdoor relative humidity ($H_a$)              \\ 
& Solar irradiance ($p_{si}$)                    \Bstrut\\
\hline
\end{tabular}%
}
\caption{Data categories and types used by the proposed approach.}
\label{tab_data_items}
\end{table}

According to the selected type of EPI and noisy factors, the data used by the proposed approach comprise AC power consumption and AC operation settings of the benchmarked rooms, indoor temperature, areas of the rooms, and ambient weather conditions. The included types of each data category are summarized in Table~\ref{tab_data_items}. All the listed data types, except for the room area, are in the form of time series, namely, sequences of time-stamped values. The sampling rate does not need to be very high as long as it can reveal the change of AC operation settings (e.g. from AC on to AC off). The required data are common and easy to collect without the need for major system modification, sophisticated sensors, or tensive manual collection.

Data preprocessing, as the first step of the benchmarking procedure, prepares the raw time-series data for later data analysis. Since the selected EPI of each benchmarked room is the average power consumed by the AC system in a given operation segment, historical AC operation segments are first extracted from the time series. For each segment, the feature set is computed, including values of the EPI (i.e. $\bar{p}_{ac}$) and all the segment-wise noisy factors (i.e. $\bar{T}_a$, $\bar{H}_a$, $\bar{p}_{si}$, $T_{ri}$, $\bar{T}_r$, $t_{seg}$, and $\bar{T}_{set}$).

Filtering of invalid data is conducted by two levels: segment-wise and room-wise.
For each operation segment, first, the duration (i.e. $t_{seg}$) is required to be within reasonable range ($\lbrack t_{seg\_min},t_{seg\_max} \rbrack$) such that the selected segments represent the regular behaviors. Second, the values of other noisy factors of the segment must be available and within physically reasonal ranges (e.g. $0<\bar{H}_a<100\%$).
\nomenclature[V]{$t_{seg\_min}$}{minimum duration required for a valid operation segment
\nomunit{$\mathrm{s}$}}
\nomenclature[V]{$t_{seg\_max}$}{maximum duration required for a valid operation segment
\nomunit{$\mathrm{s}$}}

In order to be successfully benchmarked, each room is required to have enough amount of historical data such that its AC energy performance can be well represented. Since the EPI is computed per operation segment, the amount of historical data is measured by the number of valid operation segments ($n_{seg}$), and its minimum required value is notated as $n_{seg\_min}$. The value of $n_{seg\_min}$ should be designed to balance between the chance a room is qualified and the representative level of its historical data, based on the average $n_{seg}$ over all benchmarked rooms.
\nomenclature[V]{$n_{seg}$}{number of historical operation segments available for a room}
\nomenclature[V]{$n_{seg\_min}$}{minimum value of $n_{seg}$ required for a valid room}

\subsection{Step 2: Predictive model construction for EPI computation}
\label{sec_pred_model_construct}

As summarized previously, the first task of the benchmarking procedure is to compute representative EPI value for each entity. When benchmarking the AC energy performance of residential rooms, in the historical data of each room, the EPI values (i.e. $\bar{p}_{ac}$) can be very diverse between operation segments due to different values of noisy factors. Therefore, it is difficult to obtain a representative EPI value by simply selecting from the historical values. Moreover, computing the representative EPI as the typical value (e.g. the mean, median, etc.) of the historical EPI values is not applicable because one cannot find the corresponding noisy factor values.

To overcome this challenge, a predictive model $F(\cdot)$ is constructed for every benchmarked room with the EPI (i.e. $\bar{p}_{ac}$) as the dependent variable and the segment-wise noisy factors (i.e. $\bar{T}_a$, $\bar{H}_a$, $\bar{p}_{si}$, $T_{ri}$, $\bar{T}_r$, $t_{seg}$, and $\bar{T}_{set}$) as the independent variables. The relation can then be expressed as:
\begin{equation}
\bar{p}_{ac} = F(\bar{T}_a, \bar{H}_a, \bar{p}_{si}, T_{ri}, \bar{T}_r, t_{seg}, \bar{T}_{set})+\varepsilon
\label{equa_pred_model}
\end{equation}
where $\varepsilon$ is the residual (error) of the predictive model.
\nomenclature[V]{$\varepsilon$}{absolute residual (error) of the predictive model $F(\cdot)$
    \nomunit{$\mathrm{W}$}}

The predictive model $F(\cdot)$ is required to represent the overall historical AC energy performance of the corresponding room, instead of forecasting its future performance. By inputting a certain set of noisy factor values into the trained predictive model, one can obtain a predicted $\bar{p}_{ac}$ value ($\hat{p}_{ac}$), which represents how the AC system would perform historically with the particular condition (i.e. noisy factor values). The $\hat{p}_{ac}$) is then a representative EPI value of the corresponding room, which can be used in the comparison stage.
As a result, the predictive model should be constructed using all the historical data of the room.

Moreover, constructing the predictive model not only helps to compute the representative EPI but also eases the process of computing the benchmark for each room. When the same set of noisy factor values are input into the predictive models of different rooms (with similar room area), the predicted $\bar{p}_{ac}$ values can be used to compare the AC performance of the corresponding rooms. A benchmark can then be selected from these predicted $\bar{p}_{ac}$ values.
\nomenclature[V]{$\hat{p}_{ac}$}{predicted average power consumed by the AC system
    \nomunit{$\mathrm{W}$}}

The steps of constructing the predictive model for each room are depicted in Fig.~\ref{fig_benchmarking_framework}. Before training or deploying the model, the input data are normalized. The rest of this subsection is focused on the selection of the best regression model structure and the modeling of the residuals of the trained predictive models.

\subsubsection{Selection of the best model structure}

Since the purpose of the predictive model is to accurately predict $\bar{p}_{ac}$ given certain values of the segment-wise noity factors, the model is essentially a regression model.
When constructing a regression model, different combinations of model types and settings of hyper-parameters result in different model accuracies~\cite{gunay2017inverse}. As a result, for a particular regression task, the most suitable model type should be selected, and its hyper-parameters need to be tuned.
In the case of predicting $\bar{p}_{ac}$, because every room has its own historical dataset with specific characteristics, there is a need to select the best combination of model type and settings of key hyper-parameters for every room.
To achieve this, the different combinations of model types and settings of key hyper-parameters are tested and the best-performing one is adopted.
Because the key hyper-parameters affect the concrete structure of the regression model, the combination of model type and settings of key hyper-parameters is referred to as the model structure in the following text.

In order to propose a list of model structures to test and select from, four commonly used regression model types are considered, including linear regression (LR), support vector regression (SVR), regression tree (RT), and artificial neural network (ANN).
For each model type, different settings of key hyper-parameters are selected to form up model structures based on the characteristics of the particular model type.
In general, the more settings are tested, the better performance the model has. However, as will be shown in Section~\ref{sec_case_study_pred_model}, model types have a larger impact on the performance than settings of the hyper-parameters, and the improvement made by changing the settings of the hyper-parameters is limited. Moreover, because the model residual will be properly handled in Section~\ref{sec_model_residual}, it is not necessary to increase the model accuracy by few percentages at the cost of testing all settings of the hyper-parameters. As a result, for each model type, only a few most possible settings of the key hyper-parameters are selected and tested, as listed below.
\nomenclature[A]{LR}{linear regression}
\nomenclature[A]{SVR}{support vector regression}
\nomenclature[A]{RT}{regression tree}
\nomenclature[A]{ANN}{artificial neural network}

For LR, besides its normal version (LR-normal), robust LR (LR-robust), with Bisquare as the residual weighting function, is adopted as a candidate to relieve the influences of outliers.
For SVR, both linear and non-linear kernels are considered, including SVR with linear kernel (SVR-LKN) and with Gaussian kernel (SVR-GKN).
For RT, pruning helps to reduce the complexity as well as to avoid over-fitting, so two pruned RT structures, namely RT pruned maximumly by 3 levels (RT-PB3L) and RT pruned maximumly by 5 levels (RT-PB5L) are included in the model structure list, besides the original RT with full levels (RT-full).
As for ANN, six variations are considered, namely L1S5 (one hidden layer with five hidden nodes), L1S10 (one hidden layer with ten hidden nodes), L1S15 (one hidden layer with 15 hidden nodes), L2S5 (two hidden layer with five hidden nodes), L2S10 (two hidden layer with ten hidden nodes), and L2S15 (two hidden layers with 15 hidden nodes in each layer).
As a result, considering both the model types and different settings of their key hyper-parameters, 13 model structures are included in total, namely LR-normal, LR-robust, SVR-LKN, SVR-GKN, RT-full, RT-PB3L, RT-PB5L, ANN-L1S5, ANN-L1S10, ANN-L1S15, ANN-L2S5, ANN-L2S10, and ANN-L2S15.

For some of the above model structures, there could be other hyper-parameters to settle in their actual implementations, which do not change the model structure but have an impact on the training process. Since the focus of this study is on benchmarking instead of building the most accurate predictive model, the Statistics and Machine Learning Toolbox in MATLAB 2017b is used to implement all the model structures and handle these less-important hyper-parameters.

\begin{algorithm}[!t]
\caption{Automatic selection of the best regression model structure.}
\label{algo_model_selection}
\begin{algorithmic}[1]
\Require dateset containing $y$ (i.e. $\bar{p}_{ac}$) and $x$ (i.e. $[\bar{T}_a, \bar{H}_a, \bar{p}_{si}, T_{ri}, \bar{T}_r, t_{seg}, \bar{T}_{set}]$) of the particular room, $m$ model structures \{$\mathrm{M}_1$, $\mathrm{M}_2$, ..., $\mathrm{M}_m$\}, number of folds for CV ($K_{cv}$), number of trials of CV ($N_{cv}$)
\Ensure selected regression model structure $\mathrm{M}^*$ for the particular room
\Function{AutoSelection}{}
\For{each $\mathrm{M}_i$} \Comment{iterate over all model structures}
  \State $iTrial \gets 1$;
  \While{$iTrial <= N_{cv}$} \Comment{conduct multiple trials}
    \State Divide the whole dataset into $K_{cv}$ groups randomly;
    \State $\lbrace \mathrm{DG}_j | j=1,...,K_{cv} \rbrace \gets \text{all data groups yielded by the division}$;
    \For{each $\mathrm{DG}_j$} \Comment{conduct cross-validation}
        \State Train regression model with structure $\mathrm{M}_i$ using $\lbrace \mathrm{DG}_l | l \not= j \rbrace$;
        \State $\hat{Y}_j \gets \text{all } \hat{y} \text{ predicted by the trained model with } x \text{ in } \mathrm{DG}_j$ \text{ as}
      \State the input;
    \EndFor
    \State $\mathrm{MAPE}_i(iTrial) \gets \mathrm{MAPE} \text{ computed using } \lbrace \hat{Y}_j | j=1,...,K_{cv} \rbrace$;
    \State $iTrial \gets iTrial+1$
  \EndWhile
  \State $\overline{\mathrm{MAPE}}(i) \gets \frac{1}{N} \sum_{iTrial=1}^{N} \mathrm{MAPE}_i(iTrial)$;
\EndFor
\State $i^* = \argmin\limits_{i \in \lbrace 1,...,m \rbrace } \, \overline{\mathrm{MAPE}}(i)$
\State $\mathrm{M}^* = \mathrm{M}_{i^*}$
\EndFunction
\end{algorithmic}
\end{algorithm}

To select the best regression model structure from the candidates for each room, a specific algorithm is designed as described in Algorithm~\ref{algo_model_selection}. In the algorithm, the prediction accuracy of each model structure is tested through $k$-fold cross-validation (CV) first, and the structure with the best CV accuracy is adopted.
Here, the $k$-fold CV is conducted to evaluate the ability of each model structure to predict cases that it has not encountered in the training stage. As a result, by selecting the best model structure based on the CV accuracy, overfitting of the predictive model is minimized.
Moreover, to remove the influences of the randomness of data division on the accuracy when conducting CV, the CV is conducted multiple times with different random data division and the average accuracy of all the trials is computed and used.
The model accuracy is measured by the mean absolute percentage error (MAPE) computed as:
\begin{equation}
\mathrm{MAPE} = \frac{100\%}{n_{seg}} \sum_{seg=1}^{n_{seg}} \left| \frac{\hat{p}_{ac}-\bar{p}_{ac}}{\bar{p}_{ac}} \right|
\label{equa_mape}
\end{equation}
\nomenclature[A]{CV}{cross-validation}
\nomenclature[A]{MAPE}{mean absolute percentage error}
\nomenclature[V]{$K_{cv}$}{number of folds for cross-validatiom}
\nomenclature[V]{$N_{cv}$}{number of trials of cross-validatiom for each model structure}

Among all the 12 sub-steps in Fig.~\ref{fig_benchmarking_framework}, selection of the best regression model structure has the largest computational cost, because model training is a time-consuming process, and each model structure is tested for $K_{cv} \cdot N_{cv}$ times. The total computational time of Algorithm~\ref{algo_model_selection} ($t_{A1}$) can be estimated as:
\begin{equation}
t_{A1} = N_{rooms} \cdot K_{cv} \cdot N_{cv} \cdot \sum_{i=1}^{m} t_{train}(\mathrm{M}_i)
\label{equa_tot_train_time}
\end{equation}
The computational time of other steps except model training is omitted, because they run much faster than training a regression model.
From the equation, $t_{A1}$ is decided by the number of benchmarked rooms ($N_{rooms}$), the number of folds of CV ($K_{cv}$), number of trials of CV ($N_{cv}$), and the training time of each model structure $i$ ($t_{train}(\mathrm{M}_i)$).
Therefore, the values of $K_{cv}$ and $N_{cv}$ should be set by balancing the CV quality and the computational cost. In general, values around ten are good choices.
Moreover, to control the computational cost, the number of model structures to test in Algorithm~\ref{algo_model_selection} should be limited, especially for complex model structures such as ANN.
\nomenclature[V]{$t_{A1}$}{total computational time of Algorithm~\ref{algo_model_selection}
    \nomunit{$\mathrm{s}$}}
\nomenclature[V]{$N_{rooms}$}{number of benchmarked rooms}
\nomenclature[V]{$t_{train}(\mathrm{M}_i)$}{training time of model structure $\mathrm{M}_i$
    \nomunit{$\mathrm{s}$}}

\subsubsection{Modeling of the residual}
\label{sec_model_residual}

Once the best regression model structure is selected, the predictive model with the selected structure is then trained using all historical data of the room. Nevertheless, as expressed in Equation~\ref{equa_pred_model}, the constructed predictive model cannot be 100\% accurate and there is a residual (error) term $\varepsilon$ to handle. This residual is caused by the imperfection of the predictive model as well as the randomness of human behaviors. To make the predicted $\bar{p}_{ac}$ more accurate such that the comparison of AC energy performance between rooms can be more accurate, the residual of the predictive model needs to be modeled properly.

Since the absolute error of a regression model highly depends on the magnitude of the output, the percentage error is preferred to represent the error. As a result, instead of the absolute residual $\varepsilon$, the percentage residual $\varepsilon_p$ of the predictive model is modeled as a random variable, which is computed as:
\begin{equation}
\varepsilon_p = \frac{\hat{p}_{ac}-\bar{p}_{ac}}{\bar{p}_{ac}}
\label{equa_ep}
\end{equation}
Its distribution is estimated through kernel density estimation (KDE). The sample used for conducting the KDE is obtained by running a CV of the selected model structure. As described in Algorithm\ref{algo_model_selection}, every time a CV is conducted, a set of predicted $y$ values (i.e. $\hat{p}_{ac}$) is obtained, so the sampled residual can be computed by Equation~\ref{equa_ep}.
\nomenclature[V]{$\varepsilon_p$}{percentage residual (error) of the predictive model $F(\cdot)$}

After drawing a sample of $\varepsilon_p$ from the CV, KDE with Gaussian kernel functions is then conducted to fit the probability density function (PDF) of $\varepsilon_p$. A demonstration of this process is shown in the corresponding block of Fig.~\ref{fig_benchmarking_framework}.
In the later comparison stage, the predictive model along with the fitted distribution of its $\varepsilon_p$ is used to compute stochastic $\hat{p}_{ac}$, which serves as the representative EPI of each room.
\nomenclature[A]{KDE}{kernel density estimation}
\nomenclature[A]{PDF}{probability density function}

\subsection{Step 3: Clustering of benchmarked rooms for benchmark computation}
\label{sec_clustering}

In order to conduct the evaluation of the AC energy performance of a given room, besides the representative EPI of the room, one also needs to compute a suitable benchmark for it. As discussed above, the selected type of benchmark in this approach is the peer-performance benchmark, and noisy factor equalization is a better way to eliminate the influences of noisy factors compared with simple normalization. As a result, the benchmark for each room should be computed to represent the best (or average) AC energy performance of other rooms and share similar values of the eight noisy factors with the benchmarked room. The most accurate way to obtain such a benchmark is to compute it from EPI values of comparable rooms that share similar noisy factor values with the benchmarked room. Therefore, it is necessary to first separate the benchmarked rooms into groups in which the rooms share similar noisy factor values. To achieve this, the k-means clustering algorithm is conducted.

\subsubsection{Selection of features and number of clusters}

K-means clustering algorithm separates given data points into $k$ clusters based on their feature values in the manner of the expectation-maximization algorithm. The key inputs of the algorithm are the feature vector and the number of clusters to obtain.
\nomenclature[V]{$k$}{number of clusters to obtain in the k-means clustering}

To get clusters of comparable rooms, which share the same noisy factor values, the feature vector to use in the k-means clustering should be set as the vector of all the noisy factors, namely $A_r$, $\bar{T}_a$, $\bar{H}_a$, $\bar{p}_{si}$, $T_{ri}$, $\bar{T}_r$, $t_{seg}$, and $\bar{T}_{set}$. However, when the feature dimension is high (generally larger than three), it is difficult to visualize the clustering results and the data volume needs to be large enough to get clusters with more than one data point. One way to reduce the feature dimension is to use principal component analysis (PCA), but the correlation between the eight noisy factors is not strong enough to reduce the dimension to three or less. As a result, another method to reduce the feature dimension is needed.

Since the predictive model constructed previously is able to predict the $\bar{p}_{ac}$ of a room given a certain value set of segment-wise noisy factors (i.e. all noisy factors except room area $A_r$), any two room can be set as comparable systems by inputting same value set of segment-wise noisy factors if they have similar $A_r$. Hence, ideally, the AC systems just need to be clustered by the room area, and the rest noisy factors can be equalized by using the predictive model. However, predictive model usually performs worse on input that it has not seen during the training, so to equalize the segment-wise noisy factor values between two benchmarked rooms by the predictive model, it is required that the two rooms share a common range of historical data of each segment-wise noisy factor. Based on our observations of the real-world data as well as common sense, the segment-average temperature set point ($\bar{T}_{set}$) is the only segment-wise noisy factor that different AC systems may not share common historical data range of, because different users would prefer different room temperature settings and usually stick to the same settings. Therefore, the median of historical values of $\bar{T}_{set}$ is included in the feature vector for the k-means clustering such that the rooms within the same cluster can have common range of historical values of $\bar{T}_{set}$. In summary, the feature vector used in the k-means cluster is computed as $ \lbrack A_r, \mathrm{median}(\bar{T}_{set}) \rbrack $, where $\mathrm{median}(\cdot)$ means the median of historical values of the variable. Moreover, the feature vector is normalized before the clustering.

Regarding the choice of the number of clusters ($k$) to obtain in the k-means clustering, different values of $k$ are tested and the one with the best average Silhouette value is selected.
The Silhouette value of each clustered data point measures how close the point is to all other points in the same cluster and how far the point is from points in other clusters~\cite{rousseeuw1987silhouettes}. Therefore, the mean Silhouette value of all rooms after the k-means clustering is a good evaluation of the tested $k$ value.

\subsubsection{Uniform noisy factor values for each cluster}
\label{sec_uniform_factors}

The purpose of conducting the clustering on the benchmarked rooms is to find clusters of comparable rooms that share a similar value set of the eight noisy factors. In previous k-means clustering, in order to reduce the feature dimension, only room area $A_r$ is equalized during the clustering process. Rooms within the same cluster share common range of historical values (instead of similar values) of the other seven noisy factors (i.e. $\bar{T}_a$, $\bar{H}_a$, $\bar{p}_{si}$, $T_{ri}$, $\bar{T}_r$, $t_{seg}$, and $\bar{T}_{set}$). Therefore, the equalization of these seven noisy factors needs to be done by inputting a uniform value set (of those noisy factors) into the predictive model of each room that is in the same cluster.

The computation of the uniform values of the seven noisy factors for all rooms in the same cluster is done factor by factor, but in the same manner summarized in the following. For a particular noisy factor (one of \{$\bar{T}_a$, $\bar{H}_a$, $\bar{p}_{si}$, $T_{ri}$, $\bar{T}_r$, $t_{seg}$, $\bar{T}_{set}$\}), four different ranges of its historical values of all the rooms (in the same cluster) are compared. The narrowest range that yields an overlap between all the rooms is selected, and the overlap is computed as the common range of the noisy factor between all rooms. The center of the common range is then computed as the uniform value of the noisy factor for all rooms in the same cluster. The four different value ranges tested are:
(a) [40th percentile, 60th percentile],
(b) [25th percentile, 75th percentile],
(c) [10th percentile, 90th percentile],
(d) [Minimum, Maximum].
An example of the results of this computation process is shown in the corresponding block of Fig.~\ref{fig_benchmarking_framework}. There are six benchmarked rooms in the example cluster and the distribution of historical AC segment duration of each room is shown in the box plot. Overlap of all the distributions is found when comparing the [25th percentile, 75th percentile] value range of each distribution, as highlighted by two magenta dash lines. The common value of AC operation segment duration ($t_{seg}$) for all rooms in this cluster is thus obtained as the center of the overlap, which is highlighted by the red line.

\subsection{Step 4: Comparison and computation of benchmarking scores}
\label{sec_compare_steps}

Since the previous steps of the proposed benchmarking procedure have prepared the segment-wise historical data of each benchmarked room, predictive model of each room, and clusters of comparable rooms with uniform noisy factor values, the last step is to finally compute the EPI and benchmark for each benchmarked room using all the tools, compare the two, and output the benchmarking score for the room.
As mentioned in Step~3, the adopted benchmark for a given room is computed from the EPI values of comparable rooms that share the same noisy factor values with the room. More precisely, the benchmark is selected as the EPI value of the best-performing room among the comparable rooms. Therefore, all comparable rooms in the same cluster share the same benchmark: EPI value of the best-performing room. Hence, the final comparison analysis is conducted cluster by cluster following the steps summarized in the last part of Fig.~\ref{fig_benchmarking_framework}.

Steps~4.1 and 4.2 compute predicted EPI value ($\hat{p}_{ac}$) of each benchmarked room with the uniform noisy factor values for the corresponding cluster. First, the deterministic $\hat{p}_{ac}$ is computed by inputting the uniform value set of noisy factors into the predictive model of the room. Next, to account for the residual of the predictive model, a random sample of $\varepsilon_p$ is drawn from its distribution fitted by KDE in Step~2.3. To guarantee the representative of the sample, the sample size should be as large as possible, while considering the available computation resources. By combining the sample of $\varepsilon_p$ with the deterministic $\hat{p}_{ac}$, a sample of the stochastic $\hat{p}_{ac}$ is computed as:
\begin{equation}
\mathrm{Stochastic} \: \bar{p}_{ac} = \mathrm{Determinisitic} \: \bar{p}_{ac} \times (1+\varepsilon_p)
\end{equation}
which is the final EPI values used to represent each benchmarked room in the comparison stage.

Step~4.3 first finds the best-performing room among all rooms in the cluster based on the values of their stochastic $\hat{p}_{ac}$, namely the room with the lowest $\hat{p}_{ac}$ value. The $\hat{p}_{ac}$ value of the best-performing room is thus selected as the benchmark for all the other rooms in the same cluster. Finally, the benchmarking score of each room is computed by comparing the $\hat{p}_{ac}$ value of the room with the benchmark as the following:
\begin{equation}
\eta = \frac{\hat{p}_{ac}^{*}}{\hat{p}_{ac}}
\end{equation}
where $\hat{p}_{ac}^{*}$ is the best (i.e. minimum) $\hat{p}_{ac}$ value of all rooms in the current cluster. Moreover, since the $\hat{p}_{ac}$ values obtained from Step~4.2 are a sample of stochastic $\hat{p}_{ac}$ for each benchmarked room. Step~4.3 is conducted multiple times, which equals to the sample size of the stochastic $\hat{p}_{ac}$, such that it corrects the biases caused by the predictive models and simulates the randomness of AC energy performance of the benchmarked rooms. As a result, the finally obtained benchmarking score of each room is also a random variable and a sample of it with the same size as the stochastic $\hat{p}_{ac}$ is computed and output.
\nomenclature[V]{$\hat{p}_{ac}^{*}$}{best $\hat{p}_{ac}$ value among all rooms in a given cluster
    \nomunit{$\mathrm{W}$}}

\section{Case study}
\label{sec_case_study}

\subsection{Data collection and preprocessing}

The AC dataset used in the case study was collected by a testbed set in the staff apartment of Singapore University of Technology and Design (SUTD) during the whole year of 2016 through customized Panasonic AC systems~\cite{li2017data}. Each apartment unit is comprised of one living room and one or more bedrooms.
AC data (including power consumption and operation settings) and the indoor temperature of each room were collected by the Panasonic AC systems every 30 seconds and uploaded to the central server. The dataset of ambient weather conditions used in the case study was collected by the weather station of SUTD with the frequency as once every five minutes.

Among the studied apartment units, some were rent to different tenants throughout the year, either by the bedroom or by the entire unit. Since different tenants might have different AC usage habits, we separate the whole-year data of such rooms based on their renting periods of different tenants and consider each data portion as one individual room.

The Panasonic AC systems deployed in the apartment rooms are either single-split or multi-split AC systems.
A single-split AC system contains one outdoor compressor and one indoor air outlet unit serving one room, while a multi-split system contains one outdoor compressor and multiple indoor air outlet units serving multiple rooms. Every indoor unit of a multi-split system can be controlled individually.
For rooms sharing the same compressor, when their indoor units are operating at the same time, it is difficult to differentiate how much power is consumed by each of the rooms.
Hence, in order to benchmark rooms with multi-split AC systems, for every room, we only extract AC operation segments in which only its indoor unit is operating while others are not, such that its power consumption is known.

\begin{table}[]
\centering
\resizebox{.85\textwidth}{!}{%
\begin{tabular}{cccl}
\hline
\begin{tabular}[c]{@{}c@{}}Room\Tstrut\\ area ($\mathrm{m^2}$)\Bstrut\end{tabular} & Orientation & \begin{tabular}[c]{@{}c@{}}Number\Tstrut\\ of rooms\Bstrut\end{tabular} & \multicolumn{1}{c}{Room ID} \\
\hline
10.9 & Southeast & 2 & R32-1, R35-1 \Tstrut\Bstrut\\
11.6 & Northeast & 1 & R24-1 \Tstrut\\
 & Southwest & 3 & R29-1, R29-3, R33-1 \Bstrut\\
12.0 & Northeast & 3 & R27-1, R30-1, R38-1 \Tstrut\\
 & Southwest & 3 & R26-1, R28-1, R37-1 \Bstrut\\
12.9 & Southwest & 3 & R22-1, R31-1, R34-1 \Tstrut\Bstrut\\
15.0 & Southwest & 4 & R23-1, R23-2, R25-2, R36-3 \Tstrut\Bstrut\\
19.4 & Northeast & 3 & R17-1, R18-1, R20-1 \Tstrut\\
 & Southwest & 4 & R15-2, R16-1, R19-1, R21-1 \Bstrut\\
27.2 & Southwest & 7 & R12-1, R12-3, R12-4, R13-2, R14-1, R14-2, R14-3 \Tstrut\Bstrut\\
37.6 & Southwest & 4 & R2-1, R7-1, R8-1, R9-1 \Tstrut\Bstrut\\
42.8 & Northwest & 6 & R1-1, R3-1, R4-1, R5-1, R6-1, R10-1 \Tstrut\\
 & Southeast & 1 & R11-1 \Bstrut\\
\hline
\end{tabular}%
}
\caption{Areas and orientations of the studied rooms. The first two digits of room ID indicate the physical room and the last digit indicates different tenants.}
\label{tab_room_area_orientation}
\end{table}

After conducting the above two case-specific data preprocessing steps and Step~1, in total 44 rooms (including 6 virtual rooms due to tenant separation) are selected and used in latter benchmarking analysis.
Tn the filtering process, namely Steps~1.3 and 1.4, the minimum segment duration ($t_{seg\_min}$) is set as one hour, while the maximum duration ($t_{seg\_max}$) is set as 24 h based on the observations on the data of how long the users normally operated the AC system continuously. Moreover, the minimum number of historical segments required for a valid room ($n_{seg\_min}$) is set as 20 such that the rooms are made use of as many as possible while making sure that there are enough data to compute the representative EPI for each room.
Table~\ref{tab_room_area_orientation} shows the areas and orientations of the studied 44 rooms. From the table, areas of the studied rooms are diverse, and thus the selected dataset is suitable for testing the proposed benchmarking approach.

\subsection{Predictive model construction}
\label{sec_case_study_pred_model}

Predictive model is constructed for each studied room following the Step~2 in Fig.~\ref{fig_benchmarking_framework} (details described in Section~\ref{sec_pred_model_construct}). First, the best regression model structure is selected for each room by conducting Algorithm~\ref{algo_model_selection} with $K_{cv}$ and $N_{cv}$ both set as ten. For each room, the thirteen regression model structures are tested through CV, and the one with the best CV MAPE is selected as the adopted model structure for the room.
Afterward, the predictive model of each room (with the adopted model structure) is trained using all of its historical data.
Finally, the distributions of models' percentage residuals are properly fitted by KDE. In later analysis, the model errors are taken into account in Step~4.2 by randomly sampling from the fitted distributions with a large sample size.

\begin{figure}[]
\centering
\includegraphics[width=1.0\linewidth]{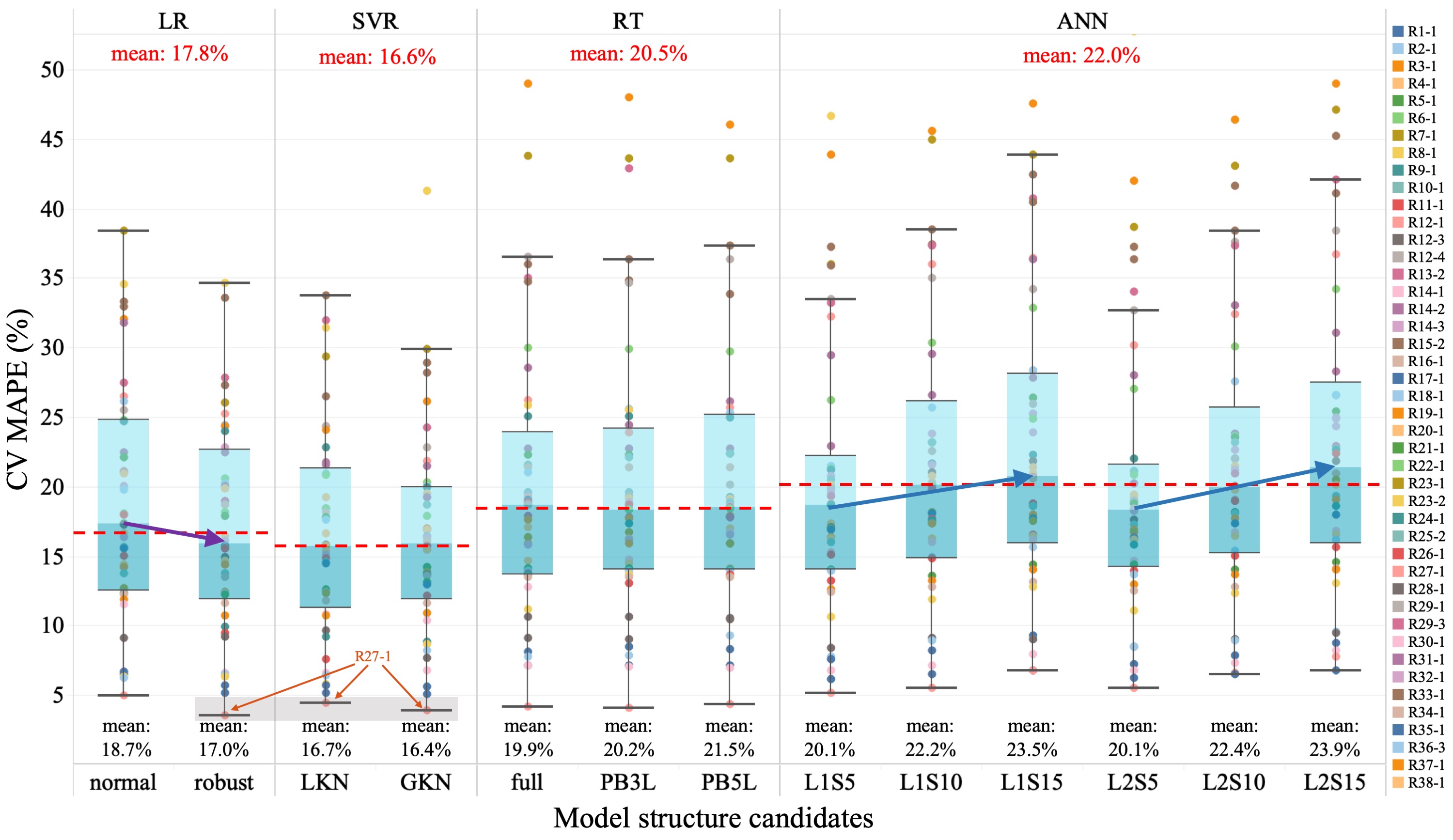}
\caption{Boxplot of CV MAPE of each model structures tested for each studied room. Columns represent different model structures (separated based on the model type). Colored dots represent the studied rooms.}
\label{fig_model_mape_boxplot}
\end{figure}

Fig.~\ref{fig_model_mape_boxplot} shows the CV MAPE of each model structure tested for each studied room, averaged over ten trials of the CV. On one hand, the performance of the same model structure for different rooms varies largely, as dots spread widely within each column. On the other hand, for the same room, different model structures yield different accuracies, as dots with the same color have different MAPE across different columns.
Among the four model types, SVR has the best accuracy with 16.6\% mean CV MAPE, while ANN performs the worst with 22\% mean CV MAPE. LR and RT are the second (17.8\%) and third (20.5\%).
The reason behind the accuracy differences between the four model types is that, for each room to benchmark, the number of available historical data points (i.e. operation segments) is limited. Therefore, the model type that can handle small datasets and minimize influences of outliers works the best. Since RT and ANN rely on the sufficiency of data amount and variety, they perform worse than LR and SVR.
Overall, the best-performing model structures are LR-robust, SVR-LKN, and SVR-GKN, with 17.0\%, 16.7\%, 16.4\% mean CV MAPE, respectively.
Moreover, the arrows in the figure show how hyper-parameters affect model performance. For LR (highlighted by the purple arrow), changing from normal version to a robust version helps decrease the error by 1.7\%. For ANN (highlighted by the blue arrows), increasing number of hidden layers does not help to improve the model performance, and increasing the number of nodes in each layer even worsens the performance. For SVR and RT, the variations in hyper-parameter settings show little impact on their performance.

\begin{table}[]
\centering
\resizebox{1.0\textwidth}{!}{%
\begin{tabular}{c|cc|cc|ccc|cccccc}
\hline
\begin{tabular}[c]{@{}c@{}}Model\Tstrut\\ structures\Bstrut\end{tabular} & \begin{tabular}[c]{@{}c@{}}LR-\\ normal\end{tabular} & \begin{tabular}[c]{@{}c@{}}LR-\\ robust\end{tabular} & \begin{tabular}[c]{@{}c@{}}SVR-\\ LKN\end{tabular} & \begin{tabular}[c]{@{}c@{}}SVR-\\ GKN\end{tabular} & \begin{tabular}[c]{@{}c@{}}RT-\\ full\end{tabular} & \begin{tabular}[c]{@{}c@{}}RT-\\ PB3L\end{tabular} & \begin{tabular}[c]{@{}c@{}}RT-\\ PB5L\end{tabular} & \begin{tabular}[c]{@{}c@{}}ANN-\\ L1S5\end{tabular} & \begin{tabular}[c]{@{}c@{}}ANN-\\ L1S10\end{tabular} & \begin{tabular}[c]{@{}c@{}}ANN-\\ L1S15\end{tabular} & \begin{tabular}[c]{@{}c@{}}ANN-\\ L2S5\end{tabular} & \begin{tabular}[c]{@{}c@{}}ANN-\\ L2S10\end{tabular} & \begin{tabular}[c]{@{}c@{}}ANN-\\ L2S15\end{tabular} \\
\hline
\begin{tabular}[c]{@{}c@{}}Number of\Tstrut\\ adopting rooms\Bstrut\end{tabular} & 6 & 6 & 9 & 18 & 0 & 2 & 1 & 0 & 1 & 0 & 1 & 0 & 0 \\
\begin{tabular}[c]{@{}c@{}}Average\Tstrut\\ training time (s)\Bstrut\end{tabular} & 0.0094 & 0.0112 & 0.0091 & 0.0083 & 0.0147 & 0.0199 & 0.0201 & 0.2212 & 0.2134 & 0.2137 & 0.2656 & 0.2671 & 0.2933 \\
\hline
\end{tabular}%
}
\caption{Number of rooms adopting the particular model structure and average training time of each model structure, computed by MATLAB 2017b in a MacBook Pro 2015 with 2.7GHz Intel Core i5 processor and 8GB memory.}
\label{tab_model_trainingtime_roomcounts}
\end{table}

From Fig.~\ref{fig_model_mape_boxplot}, SVR-GKN is the model structure with the best mean CV MAPE over all rooms, but it is not the best model structure for all rooms. For example, for room R27-1 (highlighted by gray area), the CV MAPE of LR-robust, SVR-LKN, and SVR-GKN are 3.6\%, 4.5\%, and 3.9\%, respectively, so the most suitable model structure for R27-1 is LR-robust, not SVR-GKN.
As a result, it is necessary to select the best model structure for each individual room based on the CV MAPE, instead of using the overall best structure, SVR-GKN, for all rooms.
Table~\ref{tab_model_trainingtime_roomcounts} shows how many rooms adopts each of the model structures. From the table, SVR-GKN, SVR-LKN, LR-robust, and LR-normal are the most selected model structures for the studied rooms, which is consistent with the CV MAPE shown in Fig.~\ref{fig_model_mape_boxplot}.

\begin{table}[]
\centering
\resizebox{.45\textwidth}{!}{%
\begin{tabular}{ccccc}
\hline
& Median\Tstrut\Bstrut & Mean & Min & Max \\
\hline
CV MAPE\Tstrut\Bstrut & 14.3\% & 14.9\% & 3.6\% & 29.0\% \\
CV MAE ($\mathrm{W}$)\Tstrut\Bstrut & 79.7 & 89.8 & 14.7 & 217.4 \\
CV RMSE ($\mathrm{W}$)\Tstrut\Bstrut & 123.7 & 123.4 & 21.4 & 321.5 \\
\hline
\end{tabular}%
}
\caption{Overall performance of all the constructed predictive models for the studied rooms.}
\label{tab_overall_model_performance}
\end{table}

To evaluate how accurate the constructed predictive models are in terms of predicting $\bar{p}_{ac}$, the CV MAPE values of each best-performing model structure obtained in Algorithm~\ref{algo_model_selection} are reserved and simple statistics of them are computed and shown in Table~\ref{tab_overall_model_performance}. The average CV MAPE of all best-performing model structures (for all rooms) is 14.9\% and the maximum is 29.0\%. In another word, overall the predictive model performs with the average accuracy of 85.1\% and the worst accuracy of 71.0\%.
Besides MAPE, two more measurements are computed to depict the model performance, which are mean absolute error (MAE) and root mean squared error (RMSE) computed as:
\begin{equation}
\mathrm{MAE} = \frac{1}{n_{seg}} \sum_{seg=1}^{n_{seg}} \left| \hat{p}_{ac}-\bar{p}_{ac} \right|
\end{equation}
\begin{equation}
\mathrm{RMSE} = \sqrt{ \frac{1}{n_{seg}} \sum_{seg=1}^{n_{seg}} {( \hat{p}_{ac}-\bar{p}_{ac} )}^2 }
\end{equation}
From the table, the average CV MAE and CV RMSE of the constructed predictive models (for all rooms) are 89.8 $\mathrm{W}$ and 123.4 $\mathrm{W}$, respectively. In the entire dataset, the average $\bar{p}_{ac}$ is 674.2 $\mathrm{W}$. Therefore, the average absolute residual of the constructed predictive models is 89.8 $\mathrm{W}$, which is 13.3\% of the average $\bar{p}_{ac}$. This shows the necessity of Step~2.3 of the proposed approach, namely modeling the residual of the constructed predictive model.
\nomenclature[A]{MAE}{mean absolute error}
\nomenclature[A]{RMSE}{root mean squared error}

Besides the prediction accuracy, the computational cost is another important aspect to consider when evaluating the model structures. As expressed in Equation~\ref{equa_tot_train_time}, the total computational time of Algorithm~\ref{algo_model_selection} depends on the summation of the training time of all tested model structures.
As a result, the average training time of each model structure is tested and shown in the last row of Table~\ref{tab_model_trainingtime_roomcounts}. The experiments were conducted using MATLAB 2017b in a MacBook Pro 2015, with 2.7 GHz Intel Core i5 processor and 8 GB memory.
The summation of the training time of all 13 model structures equals to 1.568 s, which means the estimated running time of Algorithm~\ref{algo_model_selection} ($t_{A1}$) is 1.92 h. Among all the model structures, six structures that are based on ANN consume the longest training time, summing up to 1.475 seconds. If removing these six structures from the structure list, $t_{A1}$ can be shortened to just 6.82 min. Given that model structures based on ANN also have the least mean CV MAPE, compared with other model types, they can be removed from the candidate list of model structures for future similar applications.
In summary, the computational time of Algorithm~\ref{algo_model_selection} can vary from several minutes to several hours. Since it is the most time-consuming step of the proposed approach, the time required to conduct the proposed approach is on the same scale. Considering that the conduction of benchmarking exercises is usually periodic and offline over a longer duration such as a week or a month, the computational time of the proposed approach is acceptable.

\subsection{Clustering of benchmarked rooms}

\begin{table}[!b]
\centering
\resizebox{0.85\textwidth}{!}{%
\begin{tabular}{cccccccccc}
\hline
$k$ & 2 & 3 & 4 & 5 & 6 & 7 & 8 & 9 & 10 \\ \hline
\begin{tabular}[c]{@{}c@{}}Silhouette\\ value\end{tabular} & 0.732 & 0.653 & 0.655 & 0.728 & 0.738 & \color{red}{0.744} & 0.701 & 0.704 & 0.713 \\ \hline
\end{tabular}%
}
\caption{Average Silhouette value over the rooms yielded by different $k$ values of the k-means clustering. Larger Silhouette value means a better separation between clusters, and thus the corresponding $k$ value is prefered. Therefore, $k=7$ is selected in the case study.}
\label{tab_silhouette_vs_k}
\end{table}

Following Step~3 in Fig.~\ref{fig_benchmarking_framework} (details described in Section~\ref{sec_clustering}), the 44 studied rooms are clustered by k-means clustering based on their room area ($A_r$) and median of their historical values of segment-wise average temperature set point ($\mathrm{median}(\bar{T}_{set})$). To select the suitable number of clusters ($k$) yielded by the k-means clustering, different $k$ values are tested and the average Silhouette value over the rooms of each $k$ value is shown in Table.~\ref{tab_silhouette_vs_k}. From the table, when $k$ equals to 7, the Silhouette value is maximum and thus the rooms are best separated. Therefore, the final number of clusters obtained is set as 7.

\begin{figure}[]
\centering
\includegraphics[width=.9\linewidth]{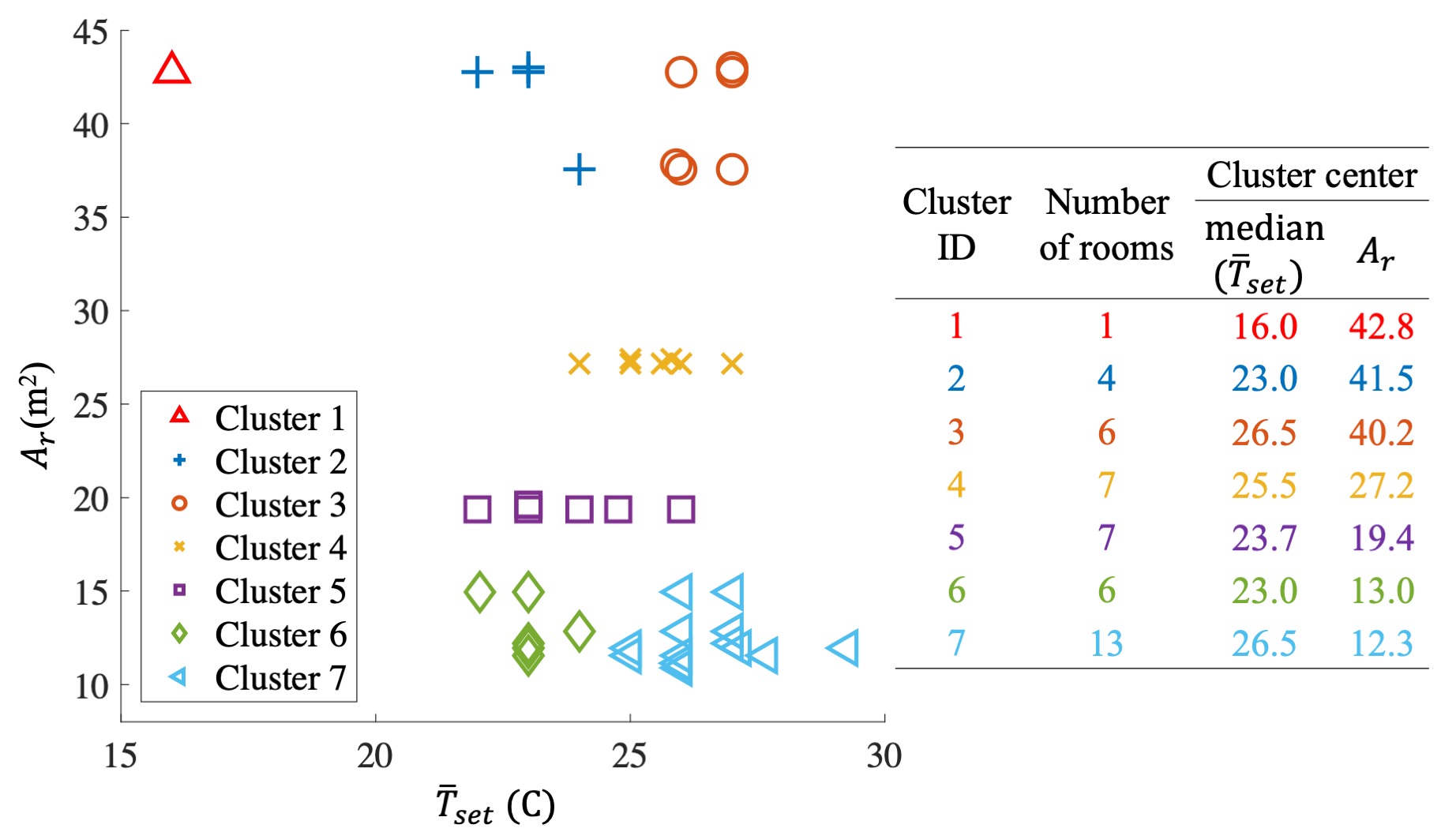}
\caption{Scatter plot of the clustered rooms, and details of each cluster. Jitter is added when two rooms have same feature values.}
\label{fig_clusters_of_rooms}
\end{figure}

After applying the k-means clustering with $k$ as 7, 7 clusters of comparable rooms are obtained as shown in the scatter plot in Fig.~\ref{fig_clusters_of_rooms}. Detailed information of each cluster is summarized in the table on the right. From the scatter plot, the studied rooms are well separated into 7 clusters based on their room area and historical temperature set point. For each of the obtained clusters, uniform values of the seven segment-wise noisy factors are computed following Step~3.2 (details described in Section~\ref{sec_uniform_factors}).

\subsection{Comparison and computation of benchmarking scores}

After obtaining the predictive model of each benchmarked room and the clusters of comparable rooms with uniform noisy factor values, the comparison between rooms within the same cluster is conducted and a benchmarking score is computed for every room, following Step~4 (details described in Section~\ref{sec_compare_steps}). The detailed results are presented in the following.

\subsubsection{Performance of the computed EPI values}
\label{sec_performace_computed_EPI}

After Steps~4.1 and 4.2, the EPI value for each room is obtained, namely the stochastic $\hat{p}_{ac}$ (predicted $\bar{p}_{ac}$) represented by a random sample. The sample size is set as 1000 here to balane between the representativeness and computational cost. For all rooms in the same cluster, they all share the similar room area ($A_r$) and their stochastic $\hat{p}_{ac}$ are computed with the uniform value set of other seven segment-wise noisy factors (i.e. $\bar{T}_a$, $\bar{H}_a$, $\bar{p}_{si}$, $T_{ri}$, $\bar{T}_r$, $t_{seg}$, and $\bar{T}_{set}$).
As a result, the influences of all the eight noisy factors have been eliminated from the EPI values for all comparable rooms in the same cluster.

\begin{figure}[!t]
\centering
\subfloat[]{\includegraphics[width=.5\linewidth]{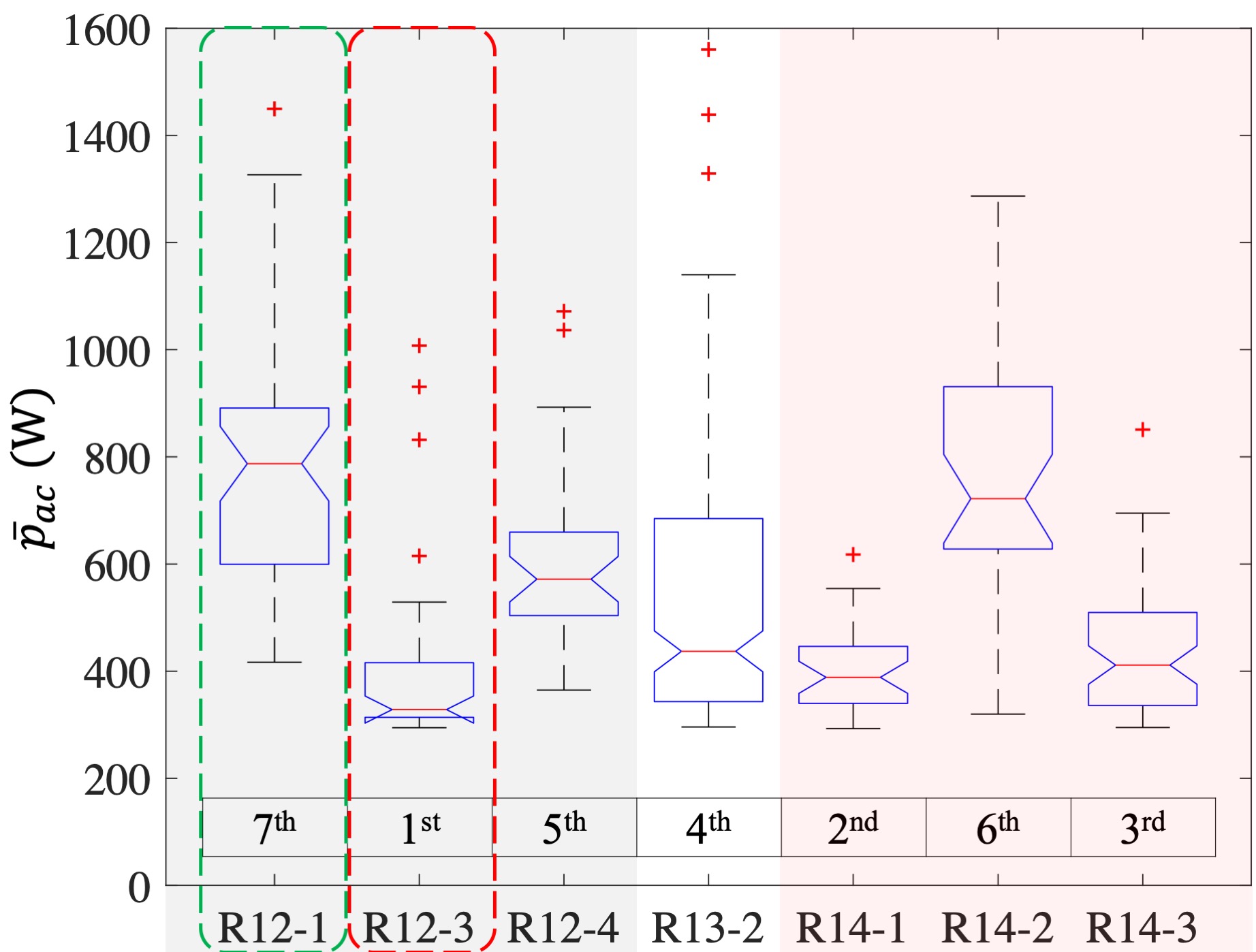}%
\label{fig_historical_p}}
\subfloat[]{\includegraphics[width=.5\linewidth]{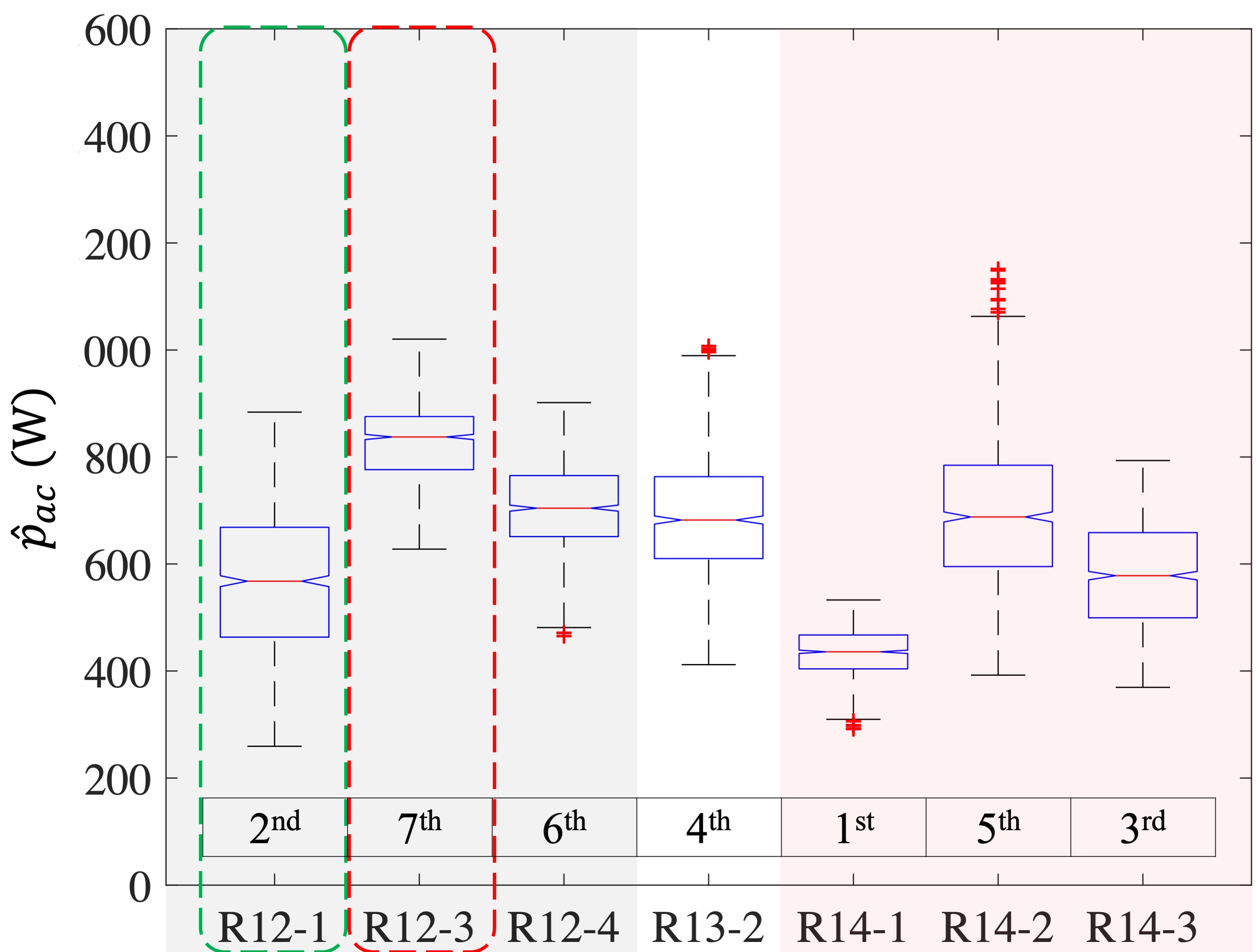}%
\label{fig_stochastic_p}}
\vfill
\subfloat[]{\includegraphics[width=.5\linewidth]{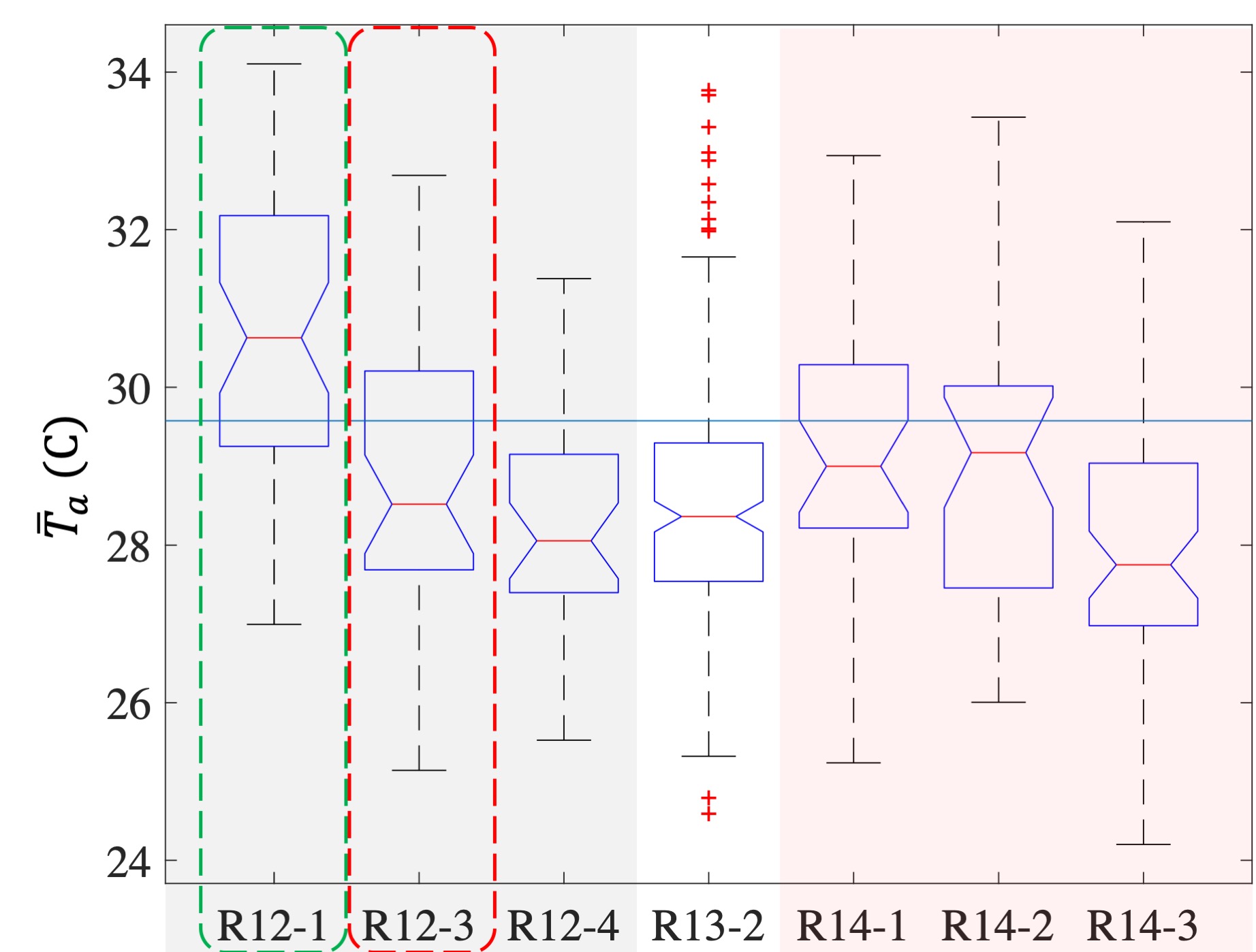}%
\label{fig_historical_Tout}}
\subfloat[]{\includegraphics[width=.5\linewidth]{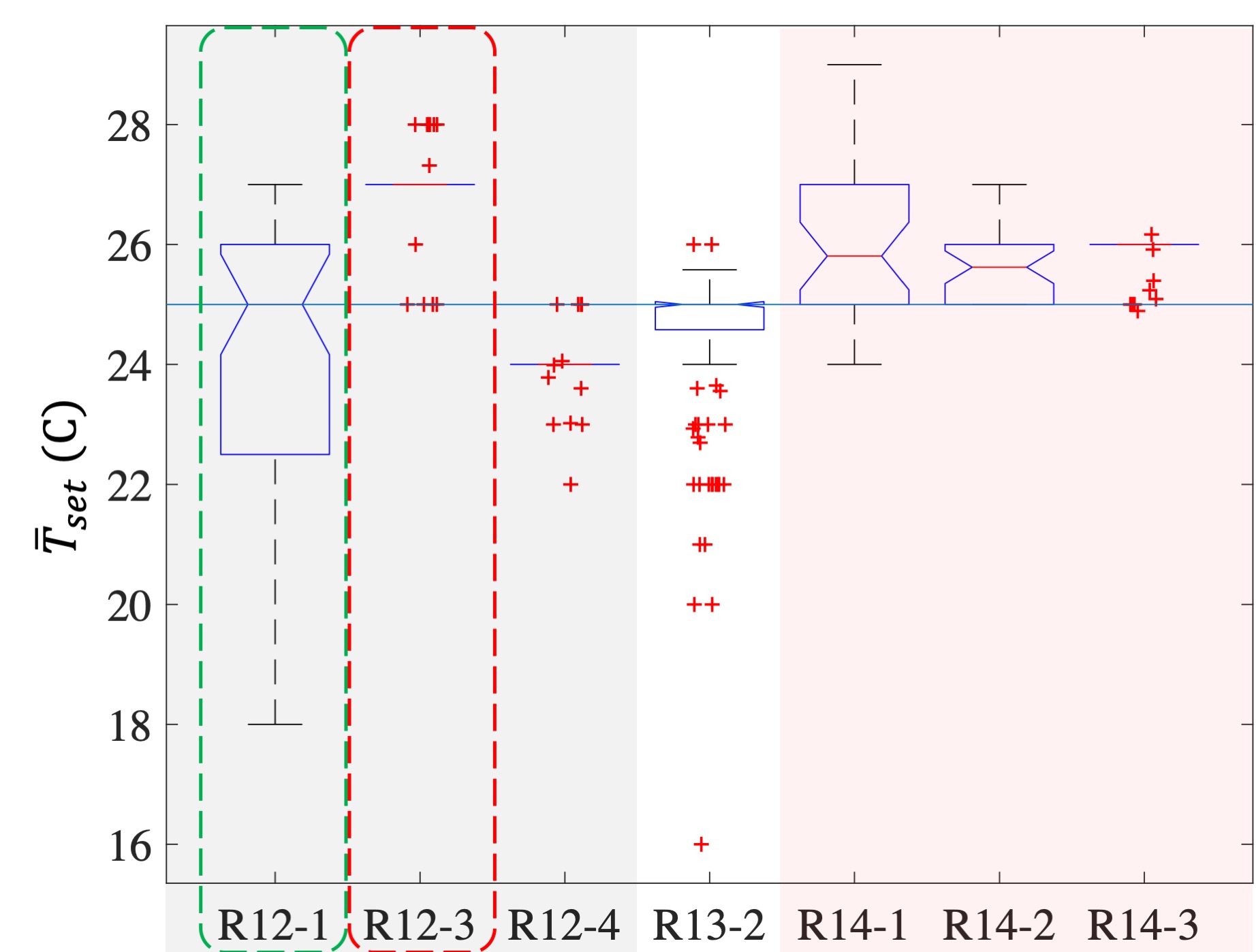}%
\label{fig_historical_Tset}}
\caption{Boxplots of (a) historical $\bar{p}_{ac}$ values, (b) stochastic $\hat{p}_{ac}$ values, (c) historical $\bar{T}_{a}$ values, and (d) historical $\bar{T}_{set}$ values of sevens rooms in Cluster 4. X-axis tick labels are room IDs. Rankings based on the power consumption (from lowest to highest) are shown in (a) and (b). Jitter is added to the outliers in (d) to avoid overlaping between dots. The cyan lines in (c) and (d) indicate the uniform values of these two factors used for this cluster.}
\label{fig_benchmarking_results_cluster4}
\end{figure}

To show the performance of the computed EPI values, in terms of the representativeness and elimination of noisy factor influences, Cluster 4 in Fig.~\ref{fig_clusters_of_rooms} is selected as the example. In Fig.~\ref{fig_benchmarking_results_cluster4}, the historical values of $\bar{p}_{ac}$, sample of stochastic $\hat{p}_{ac}$ computed by the proposed approach, and historical values of two most influential noisy factors (i.e. $\bar{T}_{a}$ and $\bar{T}_{set}$) of the seven rooms in Cluster 4 are visualized in boxplots.

From Fig.~\ref{fig_historical_p}, based on the median value of historical $\bar{p}_{ac}$, R12-1 (highlighted by the green dash line) is the most power-consuming room among the seven rooms in Cluster 4, while R12-3 (highlighted by the red dash line) consumes the least. However, if considering the historical values of their noisy factors, this case is no longer true. From Fig.~\ref{fig_historical_Tout} and Fig.~\ref{fig_historical_Tset}, historically R12-1 operates the AC with much higher ambient temperature ($\bar{T}_{a}$) and lower AC temperature set point ($\bar{T}_{set}$), compared with R12-3. Because R12-1 and R12-3 belong to the same cluster, they share similar room area (in fact, they represent the same physical room with two different tenants). Therefore, the differences in historical values of $\bar{T}_{a}$ and $\bar{T}_{set}$ could explain why R12-1 consumed more power than R12-3 in most of the historical operation segments. Moreover, there are five more noisy factors to consider.
Hence, it is necessary to eliminate the influences of all the noisy factors from the EPI values (i.e. $\bar{p}_{ac}$) such that the comparison of EPI values is fair, which is done by Steps~4.1 and 4.2 in the proposed procedure.

In Fig.~\ref{fig_stochastic_p}, the stochastic $\hat{p}_{ac}$ (predicted $\bar{p}_{ac}$) of each room computed by Steps~4.1 and 4.2 is plotted in boxplot. It is represented by distribution to account for its stochastic nature caused by the randomness of user behaviors.
Since the stochastic $\hat{p}_{ac}$ of each room is computed by predictive model and model residual distribution trained by the historical data of the room, its median value represents the average performance level of the room under the uniform condition (values of noisy factors), and its variance represents the randomness of performance.
From Fig.~\ref{fig_stochastic_p}, it is clear that after removing the influences of the seven segment-wise noisy factors, R12-1 consumes less power than R12-3 on average. In fact, R12-3 turns from the least to the most power-consuming room among the seven rooms. Comparing Fig.~\ref{fig_historical_p} and Fig.~\ref{fig_stochastic_p}, one can find that, after Steps~4.1 and 4.2, the obtained stochastic $\hat{p}_{ac}$ is a better EPI value in terms of the elimination of noisy factor influences.

In Cluster 4, there are two sets of (virtual) rooms that are from the same physical room but with different tenants, namely \{R12-1, R12-3, R12-4\} (highlighted by grey area in Fig.~\ref{fig_benchmarking_results_cluster4}) and \{R14-1, R14-2, R14-3\} (highlighted by pink area in Fig.~\ref{fig_benchmarking_results_cluster4}). By comparing the historical values of $\bar{T}_{a}$ and $\bar{T}_{set}$ of these two sets of rooms in Fig.~\ref{fig_historical_Tout} and Fig.~\ref{fig_historical_Tset}, one can see that different tenants of the same physical room had different AC usage behaviors, operating the AC in different ambient weather conditions with different AC settings. For example, tenant 3 of the physical room R14 (i.e. R14-3) normally operated the AC with a lower ambient temperature ($\bar{T}_{a}$) and a higher AC temperature set point ($\bar{T}_{set}$), compared with tenant 1 (R14-1) and tenant 2 (R14-2) of the same room. Also, tenant 3 prefers a temperature set point of 26 $\mathrm{\,^{\circ}C}$ with very few changes, while the other two tenants have wider ranges of historical set points.

Moreover, from Fig.~\ref{fig_stochastic_p}, it is obvious that different tenants of the same physical room had different levels of energy performance. For the three tenants of the same physical room R12, tenant 3 (R12-3) had the worst level of energy performance, tenant 1 (R12-1) had the best performance, and tenant 4 (R12-4) was in the middle. There is a missing tenant 2 for this room R12 because it is filtered away due to a lack of valid historical operation segments. For the three tenants of the same physical room R14, tenant 2 (R14-2) had the worst level of energy performance, tenant 1 (R14-1) had the best performance, and tenant 3 (R14-3) was in the middle.

\subsubsection{Performance of the benchmarking scores}

\begin{figure}[!t]
\centering
\includegraphics[width=\linewidth]{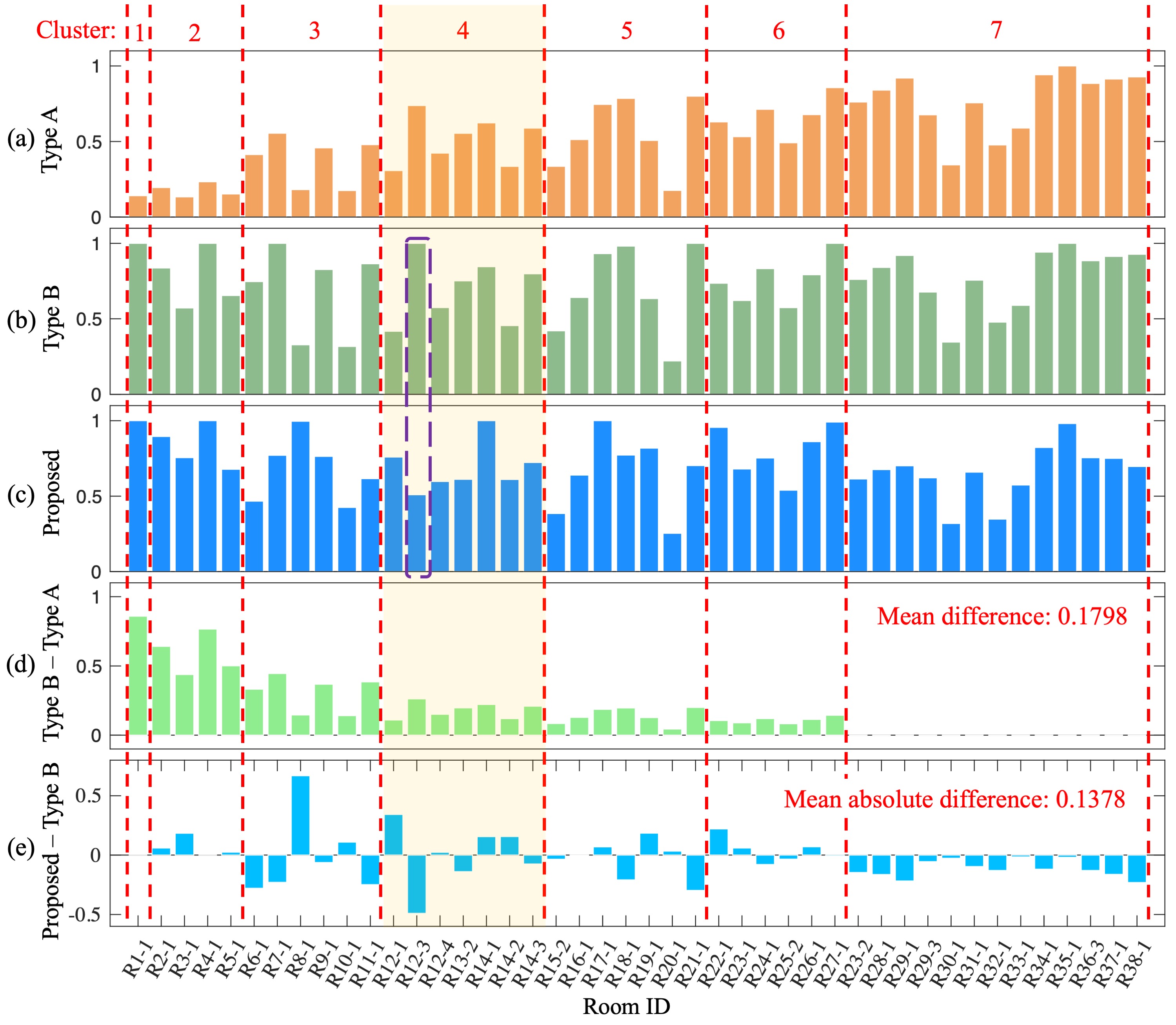}
\caption{Comparison of the benchmarking scores generated by the proposed approach and two other types of benchmarking scores of all the 44 benchmarked rooms. (a) Type A benchmarking scores, (b) Type B benchmarking scores, (c) the median values of the stochastic benchmarking scores generated by the proposed approach, (d) differences between Type B scores and Type A scores, (e) differences between scores of the proposed approach and Type B scores. The cluster ID to which each room belongs is marked at the top of the figure.}
\label{fig_diff_step_comparison}
\end{figure}

In Step 4.3 of the proposed benchmarking procedure, the stochastic $\hat{p}_{ac}$ of each room is compared with that of the best-performing room (i.e. the lowest stochastic $\hat{p}_{ac}$) in the same cluster and a benchmarking score is computed for each room as the ratio between the two. Since the EPI value is stochastic, the obtained benchmarking score is also stochastic. The median value of the obtained stochastic benchmarking score thus represents the average AC energy performance level of the room. Moreover, because the noisy factor equalization is achieved for all rooms within the same cluster, the comparison (benchmarking) is conducted within each cluster only. Hence, within each cluster, there is always one room with a stochastic benchmarking score close to one. In Fig.~\ref{fig_diff_step_comparison}c, the median values of the stochastic benchmarking scores of all benchmarked rooms are shown. To validate the overall performance of the obtained benchmarking scores, two extra types of benchmarking scores (Type A and Type B) are computed as a comparison.

Type A benchmarking score is computed as the ratio between median historical $\bar{p}_{ac}$ of each room and the lowest median historical $\bar{p}_{ac}$ among all 44 rooms, and thus none of the eight noisy factors is equalized. From Fig.~\ref{fig_diff_step_comparison}a, it is clear that there is an imbalance of benchmarking scores between clusters. From Cluster 1 to 7, the average score over each cluster is ascending. By checking the average room area of each cluster in Fig.~\ref{fig_clusters_of_rooms}, when the average room area is larger for a particular cluster, the average benchmarking score of the rooms in that cluster is poorer. Therefore, the Type A benchmarking score is highly affected by the room area and thus is not a fair benchmarking result.

Type B benchmarking score, whereas, is computed as the ratio between median historical $\bar{p}_{ac}$ of each room and the lowest median historical $\bar{p}_{ac}$ among each cluster. Since it is computed after the clustering of rooms, the influence of the room area on the score is removed.
From Fig.~\ref{fig_diff_step_comparison}d, for clusters with larger average room areas (e.g. Cluster 1 and 2), the differences between Type B scores and Type A scores are larger because these differences are used to correct the imbalance of scores between clusters existing in Type A.
From Fig.~\ref{fig_diff_step_comparison}b, based on Type B scores, there are worse-performing rooms and better-performing rooms within each cluster, which shows that Type B scores have a better balance between clusters than Type A scores.
However, since the rest seven noisy factors have not been equalized, Type B benchmarking score is still affected by those noisy factors, including $\bar{T}_a$, $\bar{H}_a$, $\bar{p}_{si}$, $T_{ri}$, $\bar{T}_r$, $t_{seg}$, and $\bar{T}_{set}$. Taking Cluster 4 as an example, from Section~\ref{sec_performace_computed_EPI} and Fig.~\ref{fig_benchmarking_results_cluster4}, the historical $\bar{p}_{ac}$ values of each room are affected by the seven segment-wise noisy factors within each cluster. Since the Type B benchmarking scores are computed from these historical $\bar{p}_{ac}$ values, the influences of these noisy factors still remain.

For the stochastic benchmarking scores generated by the proposed approach, the influences of all the eight noisy factors are properly removed. Hence the scores are further improved, as compared with Type B scores. By comparing the median stochastic benchmarking scores in Fig.~\ref{fig_diff_step_comparison}c with the Type B benchmarking scores in Fig.~\ref{fig_diff_step_comparison}b, one can find that, before and after the equalization of all noisy factors, the benchmarking scores of each particular room are different.
Fig.~\ref{fig_diff_step_comparison}e shows these differences between the scores of the proposed approach and the Type B scores. The mean absolute difference is 0.1378, which is not trivial. These differences are imposed by the proposed approach to take into account the rest seven noisy factors that Type B scores fail to do handle.
For example, in Cluster 4 (highlighted by yellow area), before the equalization, R12-3 (highlighted by the purple dash line) is the best-performing room among the seven rooms, while after the equalization, R12-3 becomes the worst-performing one. Based on the detailed analysis of Cluster 4 in Section~\ref{sec_performace_computed_EPI}, the reason behind this difference in scores is that the proposed approach takes all the eight noisy factors into account and generates fair scores with all the factors equalized.

From the above results, every key step of the proposed approach, namely Step 3 and Step 4, improves the performance of the generated benchmarking scores. Type A benchmarking score is computed without any of the key steps and thus is affected by all of the noisy factors. Type B benchmarking score is computed after Step 3, and hence the influence of room area is removed but the other seven factors still affect the scores. After conducting all the steps of the proposed approach, the obtained benchmarking scores are fairer, taking all of the noisy factors into account.

\subsection{Investigation of influences of noisy factors}

\begin{figure}[!t]
\centering
\includegraphics[width=.8\linewidth]{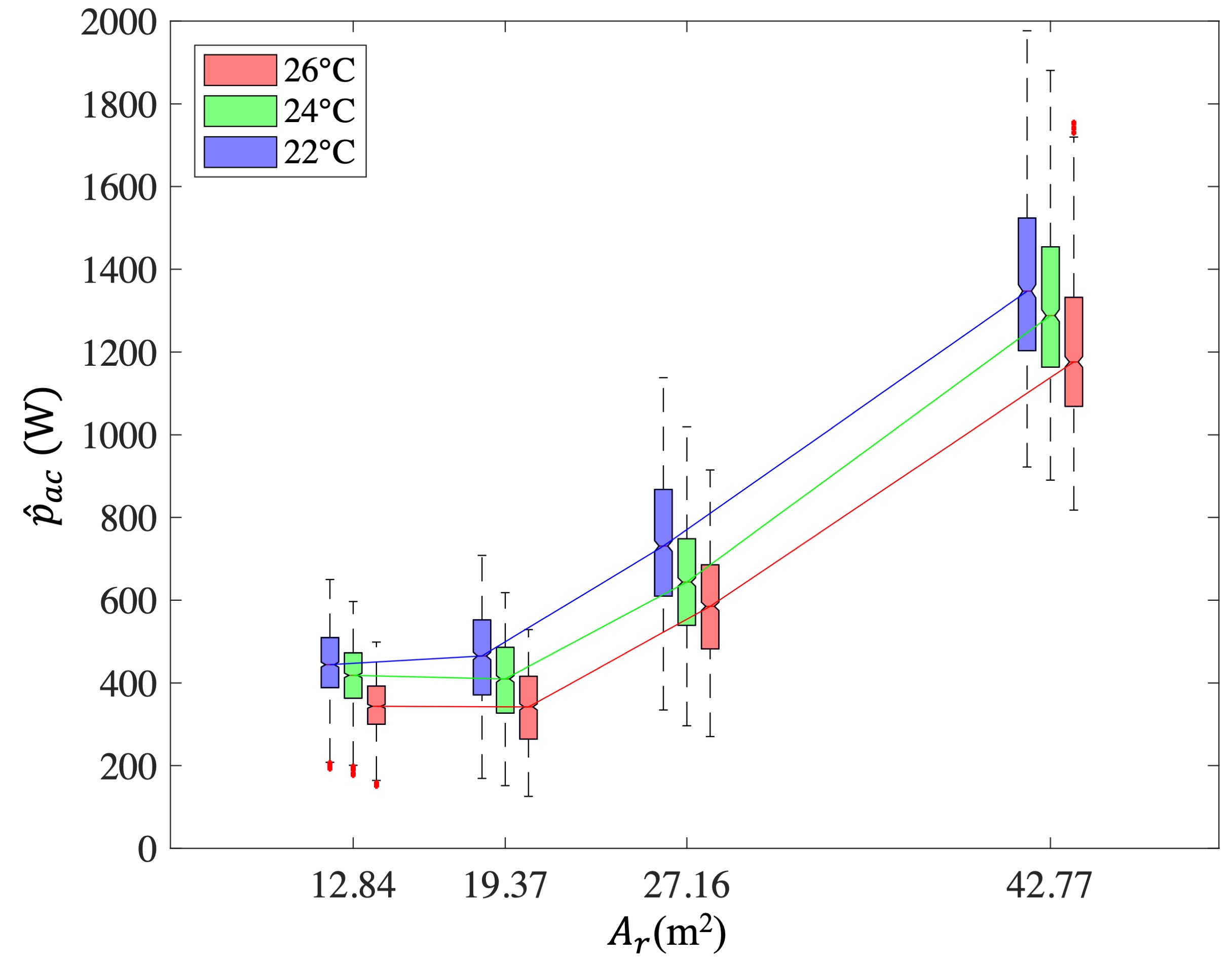}
\caption{Energy consumption (stochastic $\hat{p}_{ac}$) of the best-performing rooms with different room area and AC temperature set point. Four rooms are selected to represent the four room area levels: R22-1, R19-1, R12-1, and R4-1.}
\label{fig_power_vs_room_area}
\end{figure}

Besides generating benchmarking scores for the studied rooms, the proposed approach is also useful when investigating the influences of different noisy factors on the AC energy consumption of residential rooms. In Equation~\ref{equa_final}, the relationship between the segment-wise average power ($\bar{p}_{ac}$) and the noisy factors are expressed. However, the equation is derived from a simplified first-order thermal dynamic model of an AC-cooled room, and thus it is useful to identify the noisy factors but not precise enough to reveal the exact influences of noisy factors on $\bar{p}_{ac}$. Since the predictive model of the proposed approach is built with real-world data and is able to predict $\bar{p}_{ac}$ under a given condition (value set of noisy factors), simulation can be conducted based on the model to investigate the real-world influences of each noisy factor on the power consumption.

Here, as an example, the influences of room area ($A_r$) and AC temperature set point ($\bar{T}_{set}$) on the AC power consumption of residential rooms are investigated and results are plotted in Fig.~\ref{fig_power_vs_room_area}.
First, to investigate the influence of $A_r$, one needs to simulate the power consumption of rooms with different room area levels. From Fig.~\ref{fig_clusters_of_rooms}, there are four different room area levels of the studied rooms: around 13 $\mathrm{m^2}$, 20 $\mathrm{m^2}$, 27 $\mathrm{m^2}$, and 40 $\mathrm{m^2}$. For each room area level, one representative room should be selected.
Second, to investigate the influence of $\bar{T}_{set}$, one needs to simulate the power consumption of each room with different values of $\bar{T}_{set}$. This requires that the selected rooms have a wide range of historical values of $\bar{T}_{set}$ such that the simulation (prediction) is accurate.
Moreover, since the energy performance and other noisy factors of each room can affect the AC power consumption as well, to investigate the influences of $A_r$ and $\bar{T}_{set}$, one needs to equalize the energy performance and other noisy factors of the selected rooms. As a result, the best-performing rooms should be selected from each room area level and the simulation of all selected rooms should be conducted with other noisy factors equal to uniform values.
To fulfill the above conditions, R22-1, R19-1, R12-1, and R4-1 are selected to represent each room area level. All of them are the best or second-best performing rooms from their clusters.

From Fig.~\ref{fig_power_vs_room_area}, when the room area ($A_r$) and other noisy factors are equalized, the lower the AC temperature set point ($\bar{T}_{set}$) is, the more AC power ($\bar{p}_{ac}$) is consumed by the room. The relationship between $\bar{p}_{ac}$ and $\bar{T}_{set}$ is close to linear. When $\bar{T}_{set}$ and other noisy factors are equalized, the larger the room is, the more AC power ($\bar{p}_{ac}$) is consumed by the room. The relationship between $\bar{p}_{ac}$ and $\bar{T}_{set}$ is close to quadratic or exponential.
From the above results, it is shown that the proposed approach can be applied to investigate the influence of one particular noisy factor on the power consumption of the benchmarked rooms.

\section{Conclusion}

In this work, a data-driven benchmarking approach was proposed to fill the research gap of benchmarking the air-conditioning (AC) energy performance of residential rooms.

First, the general three steps of developing a benchmarking approach were summarized, namely identifying the benchmarking target, selecting three key elements (i.e. type of benchmark, energy performance index, and noisy factors), and designing the benchmarking procedure. Related benchmarking studies that are focused on other types of benchmarking targets were summarized and analyzed in terms of these three steps to guide the way of designing the proposed approach.

Next, the proposed approach was introduced. The benchmarking target of the approach is the AC energy performance of residential rooms. The selected type of benchmark is the peer-performance benchmark. The adopted energy performance index is the average power consumed by the AC system in one operation segment. Eight noisy factors are selected, whose influences need to be removed from the benchmarking, including room area, ambient temperature, ambient humidity, solar irradiance, initial room temperature, average room temperature, AC operation duration, and AC temperature set points. Most importantly, a benchmarking procedure particularly designed for benchmarking the AC energy performance of residential rooms was proposed. The elimination of influences of the noisy factors is achieved by equalizing the values of noisy factors, leveraging regression and clustering techniques.

Lastly, a case study was conducted with data collected from a real-world testbed. In total 44 rooms were selected and benchmarked using the proposed approach. Results showed that the constructed regression models have an average accuracy of 85.1\% in cross-validation tests in terms of predicting AC power consumption given values of noisy factors. Among the tested 13 regression model structures, support vector regression with Gaussian kernel is the overall most suitable model structure for building the regression model.
In the clustering step, 44 rooms were successfully clustered into seven clusters.
By comparing the benchmarking scores generated by the proposed approach with two sets of scores computed from historical power consumption data, we demonstrate that the proposed approach is able to eliminate the influences of room areas, weather conditions, and AC settings on the benchmarking results. Therefore, the proposed benchmarking approach is valid and fair, and it reveals the real AC energy performance of the studied rooms, which is hidden beneath the historical data. Moreover, we also demonstrated the extra usage of the proposed approach, which is to investigate the influence of each noisy factor on the AC power consumption of the benchmarked rooms in real life.

The future work of this study is threefold.
First, from Fig.~\ref{fig_diff_step_comparison}c, one can find a limitation of the proposed benchmarking approach. For an individual room that is the only member of a cluster (e.g. room R1-1 in Cluster 1), since there are no peers to compare within the same cluster, the approach cannot provide a proper score for it. This is the limitation shared by all benchmarking approaches based on the peer-performance benchmark. To overcome this limitation, an extra step can be incorporated into the overall approach to benchmark such rooms based on their previous-performance benchmarks.
Second, reasons for the performance differences between users should be explored by obtaining related data through specific surveys or more complex sensors. The extra data should include information on detailed user behaviors, room envelope status, and AC hardware status. By combining the benchmarking scores generated by the proposed approach with these data, one can provide more precise feedback to the users including recommendations about how they can improve their AC energy performance.
The third part of the future work is to evaluate the effectiveness of such feedback on the improvement of AC performance of the rooms. To achieve this, long-term data of each particular room is needed with feedback provided to the user periodically. The proposed benchmarking approach can be applied to the same room in different time periods to track performance improvement. In another word, one can treat different time periods of the same room as multiple individual rooms and apply the proposed approach. The different time periods can then be evaluated, and performance changes can be tracked.

\section*{Acknowledgments}

This work was supported in part by the National Research Foundation (NRF) Singapore and administered by Building and Construction Authority (BCA) – Green Building Innovation Cluster (GBIC) Programme Office, and in part by the SUTD-MIT International Design Centre (IDC). Any findings, conclusions, or opinions expressed in this document are those of the authors and do not necessarily reflect the views of the sponsors.

\section*{References}

\bibliography{mybibfile}

\end{document}